\documentclass[sigconf,nonacm]{acmart}   % [arXiv preprint] nonacm: 会議行・ACM著作権表示・ACM Reference Format を全て抑制

\AtBeginDocument{%
\providecommand{\BibTeX}{{%
\normalfont B\kern-0.5em{\scshape i\kern-0.25em b}\kern-0.8em\TeX}}}

\setcopyright{none}

\usepackage{enumitem}

\usepackage{xac}
\usepackage{tabularx}
\usepackage{adjustbox}
\usepackage{graphicx}
\usepackage{booktabs}
\usepackage{geometry}
\usepackage[draft,commandnameprefix=ifneeded]{changes}

\usepackage{comment}

\usepackage{ascmac}
\usepackage{fancybox}

\usepackage{hyperref}

\usepackage{listings}
\usepackage{upquote}

\usepackage[dvipsnames]{xcolor}
\usepackage[strict]{changepage}
\usepackage{framed}
\definecolor{formalshade}{rgb}{0.95,0.95,1}
\newenvironment{formal}{%
\MakeFramed{\advance\hsize-\width\FrameRestore}%
\noindent
\hspace{-4.55pt}% disable indenting first paragraph
\begin{adjustwidth}{}{7pt}%
\vspace{2pt}
\vspace{2pt}%
}{%
\vspace{2pt}
\end{adjustwidth}\endMakeFramed%
}

\definecolor{grayshade}{rgb}{0.95,0.95,0.95}
\newenvironment{prompt}{%
\MakeFramed{\advance\hsize-\width\FrameRestore}%
\noindent
\hspace{-4.55pt}% disable indenting first paragraph
\begin{adjustwidth}{}{7pt}%
\vspace{2pt}
\vspace{2pt}%
}{%
\vspace{2pt}
\end{adjustwidth}\endMakeFramed%
}

\definecolor{rqshade}{rgb}{0.93,0.96,1}
\newenvironment{researchq}{ %
\MakeFramed{\advance\hsize-\width\FrameRestore}%
\noindent
\hspace{-4.55pt}% disable indenting first paragraph
\begin{adjustwidth}{}{7pt}%
\vspace{2pt}
\vspace{2pt}%
}{%
\vspace{2pt}
\end{adjustwidth}\endMakeFramed%
}
\begin{document}
  \title{AToM CoWriter: Towards Cognitive Process–Aware Proactive Writing Support}

  %% By default, the full list of authors will be used in the page
  %% headers. Often, this list is too long, and will overlap
  %% other information printed in the page headers. This command allows
  %% the author to define a more concise list
  %% of authors' names for this purpose.
  \author{Masahiro Yoshida}
  \authornote{Both authors contributed equally to this research.}
  \orcid{0009-0000-2394-1725}
  \affiliation{%
    \institution{Sony Group Corporation}
    \city{Tokyo}
    \country{Japan}
  }
  \email{Masahiro.C.Yoshida@sony.com}

  \author{Atsuya Kobayashi}
  \authornotemark[1]
  \orcid{0009-0003-2535-8148}
  \affiliation{%
    \institution{Sony Group Corporation}
    \city{Tokyo}
    \country{Japan}
  }
  \email{Atsuya.Kobayashi@sony.com}
  
  \author{Kei Tateno}
  \orcid{0009-0000-8249-2659}
  \affiliation{%
    \institution{Sony Group Corporation}
    \city{Tokyo}
    \country{Japan}
  }
  \email{Kei.Tateno@sony.com}

  \author{Xiang ``Anthony'' Chen}
  \orcid{0000-0002-8527-1744}
  \affiliation{%
    \institution{University of California, Los Angeles}
    \city{Los Angeles}
    \state{California}
    \country{USA}
  }
  \email{xac@ucla.edu}

  \renewcommand{\shortauthors}{Yoshida et al.}

  %%
  %% The abstract is a short summary of the work to be presented in the
  %% article.
  \begin{abstract}
    Large language models can support writing, but existing tools require users to explicitly articulate prompts---particularly burdensome in creative writing, where intentions are often ambiguous.
Proactive support that infers users' needs from writing interactions could alleviate this burden, but raises two challenges: determining \textit{what} support to provide and \textit{when} to intervene.
This work focuses on the former.
We hypothesize that Flower and Hayes' cognitive process theory of writing---which characterizes writing through six cognitive processes---offers an interpretable bridge between observable writing behavior and appropriate support types.
Through a formative study and literature review, we identify 14 writing support types associated with these cognitive processes, along with characteristic interaction behaviors linked to each process. We then instantiate this framework in \textit{AToM CoWriter}, which infers support needs from writing interactions and document context.
Two within-subjects studies (N = 21) provide initial evidence that this approach improves expressiveness and idea exploration, and that incorporating interaction-derived cognitive-process cues was associated with greater engagement with proactive suggestions.
These findings suggest that cognitive processes can provide a promising basis for support selection in proactive writing systems.

  \end{abstract}

  %% CCS concepts (required by ACM).  Generate the full CCSXML block with the
  %% ACM CCS tool (https://dl.acm.org/ccs) before the camera-ready version.
  % \ccsdesc[500]{Human-centered computing~Interactive systems and tools}

  %% Keywords. The author(s) should pick words that accurately describe
  %% the work being presented. Separate the keywords with commas.

  \begin{teaserfigure}
    \includegraphics[width=\textwidth]{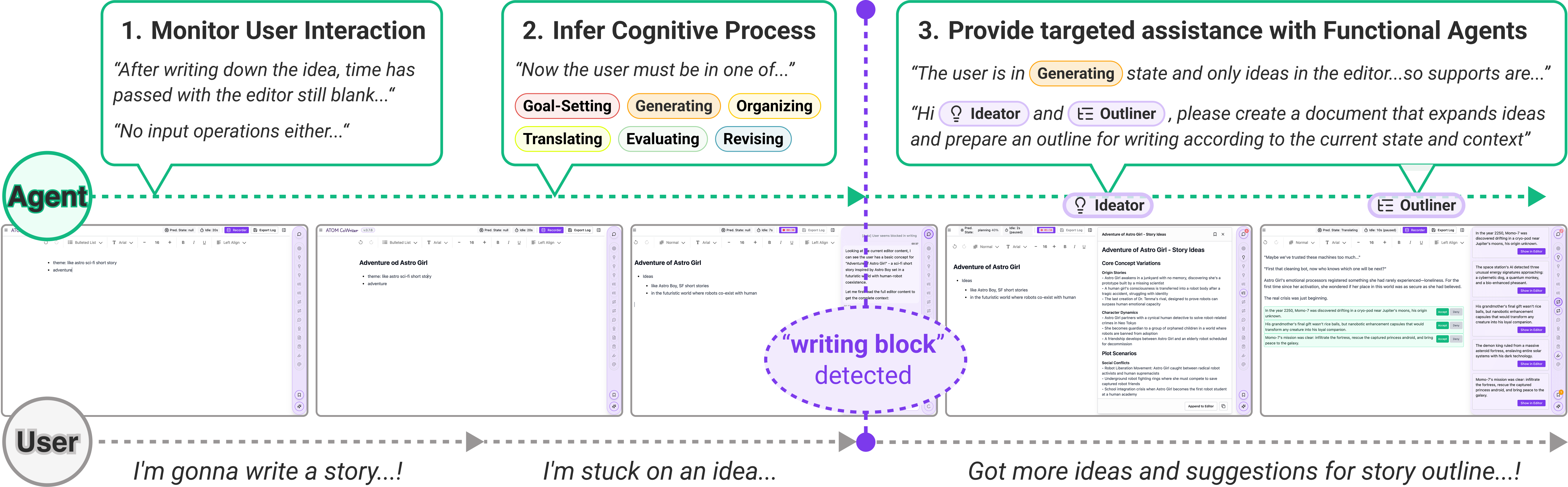}
    \caption{%
    Proactive support workflow in \textit{AToM CoWriter}.
    The system (1) continuously monitors writing interaction behaviors (keystrokes, cursor movement, editing patterns),
    (2) infers the user's current cognitive process from a summary of these interaction signals together with the current document text, using an LLM-based prediction grounded in the Flower--Hayes model~\cite{Flower1981-vp}, and
    (3) upon detecting a potential writing block, automatically activates appropriate Functional Agents to generate targeted support.
    The user can then access these suggestions to advance writing without explicit prompting.
    }
    \label{fig:teaser}
  \end{teaserfigure}

  %% This command processes the author and affiliation and title
  %% information and builds the first part of the formatted document.
  \maketitle

  \section{Introduction}
\label{sec:introduction}

Large language models (LLMs) are reshaping how people approach writing tasks.
Chat-based LLM services such as ChatGPT\footnote{ChatGPT by OpenAI, \url{https://chat.openai.com}} and Claude\footnote{Claude by Anthropic, \url{https://claude.ai}} support all phases of writing—from ideation and outlining to drafting and reviewing—through appropriately crafted prompts.
These advances highlight the potential of AI agents as collaborators that augment human writing capabilities \cite{Lin2023-op}.

However, in these user-initiated systems, writers must pause their work to reflect on their current state and challenges, and then articulate an appropriate prompt that conveys their intent and context \cite{Guo2025-ju}.
This process involves two key burdens: the cognitive load of verbalizing problems and the physical effort of prompting~\cite{Subramonyam2024-tn, Kuo2026-ps}.
As a result, prompt-driven support, while flexible, can interfere with the natural writing flow.
These burdens are particularly salient in creative writing, such as storytelling, where writers' intentions are often ambiguous and their support needs may change dynamically throughout the writing process.

In contrast, proactive writing support shifts part of this burden from the writer to the system: rather than requiring writers to specify each request explicitly, the system can infer likely support needs from naturally occurring writing interactions and offer assistance accordingly.
Such inference may draw on signals including keystrokes, mouse operations, cursor movements, editing patterns, and the evolving document \cite{Liu2026-aq}.
In the domain of proactive support, many systems have been proposed that automatically complete or continue text \cite{Bhat2023-uk, Jakesch2023-ih}. 
While highly useful, such support primarily targets the \textit{Translating} process, in which writers transform ideas into linguistic form~\cite{Flower1981-vp}.
Writing difficulties, however, also arise while generating ideas, organizing thoughts, evaluating drafts, and revising text.
Supporting these different activities requires not only detecting when assistance may be needed, but also determining what kind of assistance is appropriate at a given moment.

Designing such proactive support therefore involves two distinct challenges: determining \textit{\textbf{what}} to offer and \textit{\textbf{when}} to offer it.
This work focuses on the \textit{what} problem.
A central challenge is to connect observable writing behaviors with the types of support that writers may need at a given moment. We investigate cognitive processes as a theory-grounded bridge between these behaviors and appropriate support types.
Specifically, we ask whether observable writing interactions can provide useful cues about the writer's current cognitive process, and whether this process can in turn guide the selection of appropriate writing support.
We distinguish this question from the \textit{when} problem.
Existing proactive systems commonly use heuristic triggers such as inactivity~\cite{Buschek2021-lc, Bhat2023-uk}.
Our system adopts the same heuristic: it does not adapt intervention timing to the writer, and we leave adaptive timing to future work (Sec.~\ref{sec:discussion-timing}).
Our scope is creative story writing, where writers repeatedly move among ideation, organization, translation, evaluation, and revision, providing a suitable setting for investigating process-aware support selection.

To operationalize this approach, we conducted a formative study grounded in the cognitive process model of writing by Flower and Hayes~\cite{Flower1981-vp}, which characterizes writing as an iterative interplay of six cognitive processes and has been widely applied in the design and analysis of writing support tools~\cite{Wan2024-qq, Gero2022-ad}.
Through this study and a complementary literature review, we identified fourteen types of writing support associated with six cognitive processes, together with characteristic writing interaction behaviors related to each process.
These findings provide a theory-grounded mapping from observable writing interactions, through cognitive processes, to candidate support types (Sec.~\ref{sec:formative}).

We instantiate this framework in \textit{AToM CoWriter}, a proactive writing support system for creative writing.
AToM CoWriter continuously observes writing interactions such as typing pauses and speed, deletion and revision patterns, cursor location and movement, and scrolling, estimates the writer's current cognitive process together with document context, and uses this information to select relevant forms of assistance.
The system consists of a \textit{Main Agent}, which coordinates support selection, and fourteen \textit{Functional Agents} corresponding to the support types identified in the formative study.
When support is triggered---in the current implementation, after a brief period of typing inactivity---the Main Agent selects up to three relevant Functional Agents, which generate suggestions in the background for users to inspect at their own pace.
The name AToM (AI with Theory of Mind) is inspired by the concept of Theory of Mind—the ability to infer others' intentions, goals, and emotions \cite{Wang2024-ie, Wang2021-ba}. In AToM CoWriter, we draw on this idea to infer writers' likely cognitive processes from their writing interactions.
Suggestions are surfaced through an unobtrusive notification rather than a pop-up, and writers can additionally request support at any moment through a user-initiated, prompt-less mechanism.

We evaluated this approach through two complementary within-subjects studies on short story writing.
Study~1 ($n=11$) examined the overall experience of system-inferred support relative to conventional user-articulated, prompt-based support, providing initial evidence of benefits for expressiveness and idea exploration while also revealing tensions around timing and user control.
Study~2 ($n=10$) held the proactive interaction paradigm constant and examined whether adding interaction-derived cognitive process cues to document-based support selection affected engagement with proactive suggestions.
Interaction results showed substantially greater engagement with proactively generated suggestions when cognitive process inference was incorporated, providing evidence that behavioral process cues can contribute to support selection.

In summary, this paper makes the following contributions:
\begin{enumerate}[leftmargin=5.5mm]
    \item A theory-grounded framework for incorporating cognitive process inference into proactive writing support, connecting observable writing interactions to the six cognitive processes defined in the Flower--Hayes model and to fourteen support types identified in our study (Sec.~\ref{sec:formative}).

    % \item An instantiation of this framework in \textit{AToM CoWriter}, which uses writing interaction traces and document context to infer cognitive processes and select among multiple forms of proactive writing assistance (Sec.~\ref{sec:design}).

    \item The design and implementation of \textit{AToM CoWriter} as an instantiation of this framework, using writing interaction behaviors and document context to infer cognitive processes and provide context-appropriate proactive support (Sec.~\ref{sec:design}).

    \item Empirical evidence from two complementary user studies suggesting that system-inferred support can facilitate creative exploration, and that incorporating interaction-derived cognitive process cues can increase writers' engagement with proactively generated suggestions (Sec.~\ref{sec:result}).
\end{enumerate}

Together, these contributions suggest that cognitive processes can serve as a useful intermediate representation for determining \textit{what} support to provide in proactive writing systems.
  \section{Related Work}
\label{sec:related-work}

\subsection{Proactive AI: What and When to Offer}
\label{sec:related-proactive}

Systems that respond only to explicit requests require users to first recognize their own difficulty and then articulate it, which itself imposes cognitive and interaction costs~\cite{Subramonyam2024-tn,Kuo2026-ps}.
Proactive and mixed-initiative interaction instead shifts part of the initiative from users to the system, and has been extensively studied to reduce these costs~\cite{Horvitz1999-jy,Allen1999-jl}, with a longstanding challenge being how to balance system initiative with user control~\cite{Chu-Carroll2000-vb}.
Poorly timed or misaligned interventions can increase interaction costs or undermine users' sense of agency~\cite{Horvitz1999-jy,Clark2018-iz}, making proactivity particularly challenging in open-ended activities such as writing.

We characterize proactive support in terms of two key design questions: \textit{when} assistance should be offered and \textit{what} assistance should be provided.
Regarding the \textit{when} question, existing systems have used a range of triggering mechanisms, including real-time intervention after each input~\cite{Chang2015-gr,Arnold2020-xk,Tsai2020-hf}, activation upon inactivity~\cite{Buschek2021-lc,Jakesch2023-ih,Bhat2023-uk}, and threshold-based triggers on editing activity~\cite{Lin2024-it}.
More recent work on proactive agents has explored adaptive approaches to intervention timing, including personalized proactivity, modeling the costs of unnecessary assistance, and timing-aware decision policies~\cite{Kaur2026-sk,Steyvers2025-kj,Ding2026-hm,Liu2025-ch}.

The \textit{what} question concerns selecting an appropriate form of assistance once support is considered useful.
Many proactive systems are designed around a particular form of assistance~\cite{Buschek2021-lc,Tsai2020-hf}, while some systems adaptively select among multiple support types.
For example, Lin et al.\ used a Multi-Armed Bandit algorithm to select among three support types~\cite{Lin2024-it}, and Chen et al.\ proposed a proactive programming assistant that uses an LLM to predict which of eight support categories is required~\cite{Chen2025-km}.
% In proactive writing support specifically, the assistance provided has most often been text continuation or completion~\cite{Bhat2023-uk,Jakesch2023-ih}, which primarily helps writers turn already-formed ideas into linguistic form.
% Writing difficulties, however, also arise while generating ideas, organizing them, evaluating a draft, and revising text, and these activities remain comparatively unsupported.

Our work focuses primarily on this \textit{what-to-offer} problem.
Support selection is particularly relevant to writing because writers iteratively move among qualitatively different cognitive processes, and the forms of assistance that are useful can differ across these processes.
The following subsection reviews this process-oriented view of writing and prior work relating writing processes to support needs.
For the \textit{when} question, we adopt a pragmatic combination of inactivity-based triggering and a user-initiated mechanism, so that the selection of support can be investigated without simultaneously introducing a new timing mechanism.

\subsection{Cognitive Processes and Writing Support}
\label{sec:related-cognitive}

Writing is not a single, linear activity but an iterative process comprising qualitatively different forms of cognitive work.
The Cognitive Process Theory of Writing by Flower and Hayes~\cite{Flower1981-vp} decomposes writing into six cognitive processes: Generating, Organizing, and Goal Setting, which together constitute Planning; Translating; and Evaluating and Revising, which together constitute Reviewing.
Later models have extended this account to accommodate additional factors such as motivation, working memory, and the social context of writing~\cite{Hayes2012-nv,Becker2006-qd}.
The process-oriented perspective has likewise been employed to analyze and design writing support systems~\cite{Gero2022-ad,Beauvais2011-de,Wan2024-qq}.

Prior work suggests that different stages or processes of writing call for different forms of assistance.
Chakrabarty et al.~\cite{Chakrabarty2024-pg} analyzed LLM-assisted writing through the Flower--Hayes model and found that the value of LLM support differed across cognitive processes.
G\"oldi et al.~\cite{Goldi2024-xp} similarly demonstrated the importance of aligning intelligent writing support with relevant cognitive processes.
More broadly, design spaces of intelligent writing assistants have organized prior systems according to the stages of writing that they support~\cite{Lee2024-hb,Gero2022-ad}, and recent work has investigated how specific writing tools correspond to writers' needs and writing activities~\cite{Zhao2025-ea}.

Together, these studies suggest that writers' support needs vary with the processes in which they are engaged.
However, as noted above, existing systems generally require writers to recognize and articulate those needs themselves.
Our work therefore uses cognitive processes as a theory-grounded intermediate layer between observable writing behavior and support selection: rather than treating process identification as an end in itself, we investigate whether it can help determine \textit{what} support a proactive writing system should provide.
% This raises a further question of whether such processes can be inferred from writers' naturally occurring interactions.

\subsection{Inferring User States from Interaction Traces}
\label{sec:related-user-modeling}

Observable interaction traces have long been used as signals for modeling users' activities, abilities, and task states~\cite{Gajos2004-wc,Cockburn2007-at,Liu2026-aq}.
In writing, such traces provide particularly rich signals for understanding and modeling writers' ongoing cognitive processes.
A substantial literature has analyzed keystroke dynamics and temporal patterns in writing, associating pauses, revision behavior, cursor activity, and other interaction measures with different writing processes and cognitive demands~\cite{Baaijen2012-my,Guo2018-us,Zhang2025-mt,Delgado-Osorio2025-ps}.
Some studies have explicitly attempted to infer phases or cognitive processes of writing from keystroke logs, textual context, or combinations of behavioral features~\cite{Conijn2024-nf,Wang2025-nr}.
Mouse and cursor behavior have likewise been studied as signals of users' cognitive activity and workload~\cite{Arshad2015-bf,Zhang2024-ln}.

Related approaches have recently appeared in AI-assisted programming, where systems observe interactions with code editors to infer whether and how assistance may be useful.
For example, proactive programming assistants have used editing activity and task context to determine potential assistance~\cite{Pu2025-zg,Mozannar2024-nc}.
These studies demonstrate how ongoing interaction can provide signals for inferring task-relevant user states and needs without relying solely on explicit commands.

Our work builds on this tradition of interaction-based user modeling in the domain of writing.
Rather than directly mapping low-level behavioral signals to a particular intervention, we use interaction traces to estimate the writer's current cognitive process.
The inferred process then structures the selection among multiple support types, connecting interaction-based user modeling with a theory of writing and providing an interpretable basis for proactive support selection.
% How the selected support is then surfaced to writers is a separate design question, which prior work has addressed through a variety of interaction paradigms.

\subsection{Interaction Paradigms for AI Writing Support}
\label{sec:related-writing-interaction}

AI-assisted writing systems have explored a broad range of interaction paradigms for supporting writers.
Existing systems provide functions such as text continuation, rewriting, critique, summarization, and idea generation through inline suggestions, dedicated interfaces, or conversational interaction~\cite{Bhat2023-uk,Jakesch2023-ih,Lee2024-hb,Chung2022-nl,Mirowski2023-qe,Bao2022-nv,Talaei2025-tg,Zhang2025-dl,Yao2026-re,Dang2022-te,Laban2024-ya}.
While chat-based LLMs provide flexible access to many such capabilities, obtaining support through them depends on the writer's own articulation of the problem, with the costs described above.

Recent research has therefore explored interaction paradigms that go beyond conventional prompt-based chat.
For example, \textit{Texterial} treats text as directly manipulable material for LLM-mediated writing~\cite{J-Shen2026-ut}, while Visual Story-Writing supports story development through manipulation of visual representations~\cite{Masson2025-kf}.
\textit{Polymind} uses parallel visual diagramming and microtasks to support prewriting~\cite{Wan2025-nl}, and \textit{Narrix} supports story writing by allowing writers to remix narrative strategies from examples~\cite{Zhang2026-xx}.
Other work has examined how writers integrate LLMs into prewriting and creative practice~\cite{Wan2024-at,Guo2025-ju}, illustrating that useful AI assistance can take substantially different forms across the writing process.
Beyond the utility of the support itself, interaction design also shapes writers' sense of agency and ownership, and prior studies have documented tensions between AI assistance, authorial control, and perceived ownership~\cite{Carrera2026-eh,Draxler2024-sa,Biermann2022-jk,Yoshida2026-bg}.

Together, these studies illustrate substantial diversity in both the forms of writing support and the ways in which writers interact with AI.
Our work is complementary to efforts that develop new interaction techniques for co-writing.
Rather than proposing a single interaction paradigm intended to accommodate all writing activities, we focus on an upstream problem: determining \textit{what} support is likely to be relevant to the writer's current process without requiring the writer to articulate that need explicitly.
% AToM CoWriter provides one concrete interaction design through which we instantiate and evaluate this process-aware support-selection approach.

  %% ============================================================================
%% [arXiv preprint] Formative Study restored as a standalone section.
%% ============================================================================

\section{Formative Study}
\label{sec:formative}
\label{sec:design-formative} %% preserved for cross-references (e.g., Sec.~4 User Evaluation)

This formative study aimed to deepen our understanding of the relationships among writers' interaction behaviors, momentary cognitive processes, and the support they require at each moment, as a foundation for designing proactive writing support.
In this study, we use the term \textit{cognitive process} to refer to the \textit{momentary cognitive process} that is most prominent at a given moment during writing, as characterized in the Flower and Hayes model~\cite{Flower1981-vp}. Unless otherwise noted, all mentions of cognitive processes in the remainder of the paper follow this definition.

To structure this investigation, we adopted the Cognitive Process Model of Writing proposed by Flower and Hayes~\cite{Flower1981-vp}, which provides a well-established framework linking observable writing interaction behaviors to the underlying cognitive processes.
To collect empirical data, we developed a custom data-logging editor grounded in this framework. Participants used the tool to write short stories, during which we recorded both detailed writing interaction logs and self-reported cognitive processes whenever they felt ``blocked.''

Building on these foundations, this section addresses the following three research questions (RQs):

\begin{researchq}
\begin{description}[leftmargin=0em, labelsep=0em, style=nextline]
  \item[\textbf{RQ-F1.}] What relationships exist between users' cognitive processes and the types of support they require? (Sec.~\ref{sec:formative-rel-cog-sup})
  \item[\textbf{RQ-F2.}] How are users' observable writing interaction behaviors related to their cognitive processes during writing blocks? (Sec.~\ref{sec:formative-rel-behav-cog})
  \item[\textbf{RQ-F3.}] Can users' cognitive processes be predicted from their behavioral data? (Sec.~\ref{sec:formative-cog-pred})
\end{description}
\end{researchq}

\subsection{Data Logging Editor}
\label{sec:formative-tool}

\begin{figure*}[ht]
    \centering
    \includegraphics[width=\textwidth]{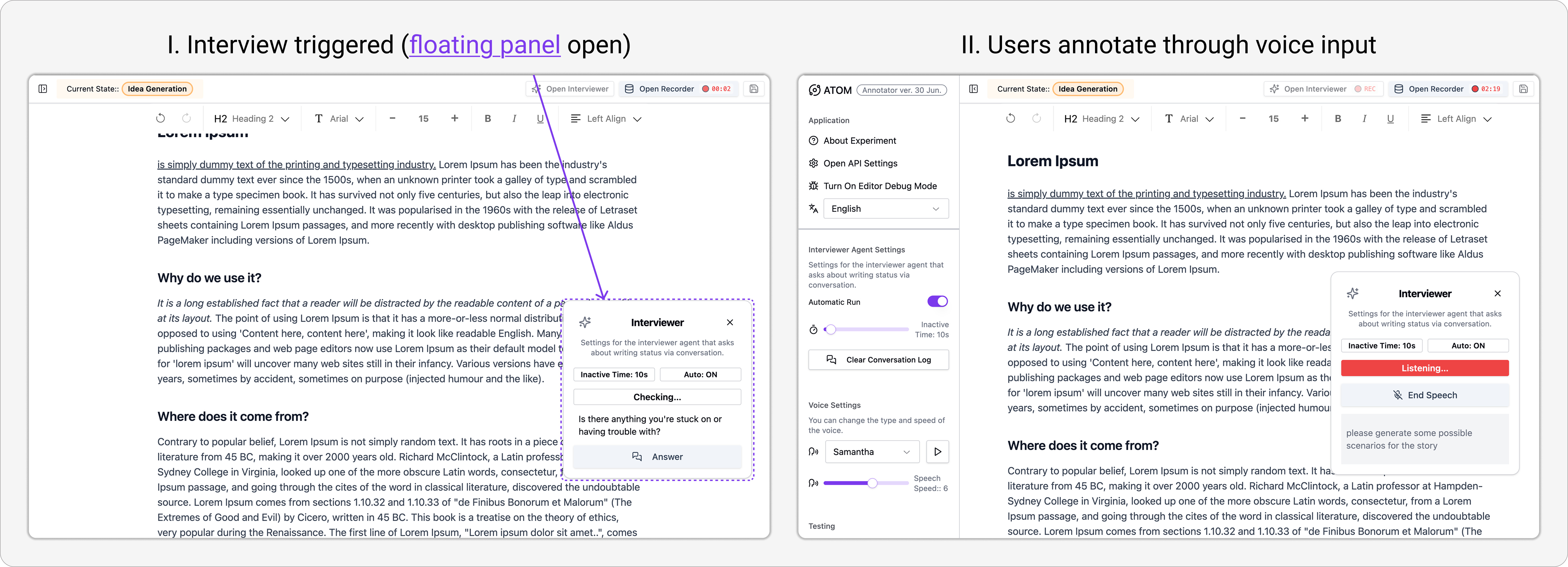}
    \caption{The data-logging editor used in the formative study. It collects cognitive process annotations via voice input to minimize disruption during writing.}
    \label{fig:annotation_tool_appendix}
\end{figure*}

The developed data-logging editor was a lightweight rich-text editor augmented with logging and annotation capabilities for later analysis (Fig.~\ref{fig:annotation_tool_appendix}). It recorded two types of data: (1) the \textit{Writing Interaction Log}, which captures users' interactions with the editor during writing; and (2) the \textit{Self-Report Log}, which records users' reflections when they encounter a writing block.
The Writing Interaction Log automatically tracks various writing-related activities, including keystrokes, mouse operations, cursor positions, and text content.
The Self-Report Log allows users to annotate their current process and thoughts whenever they feel blocked or in need of support. In other words, users primarily continue writing as usual but are encouraged to provide annotations at moments when they experience a block in their writing process.

\subsubsection{Self-Report Procedure}

\begin{figure*}[ht]
    \centering
    \includegraphics[width=\textwidth]{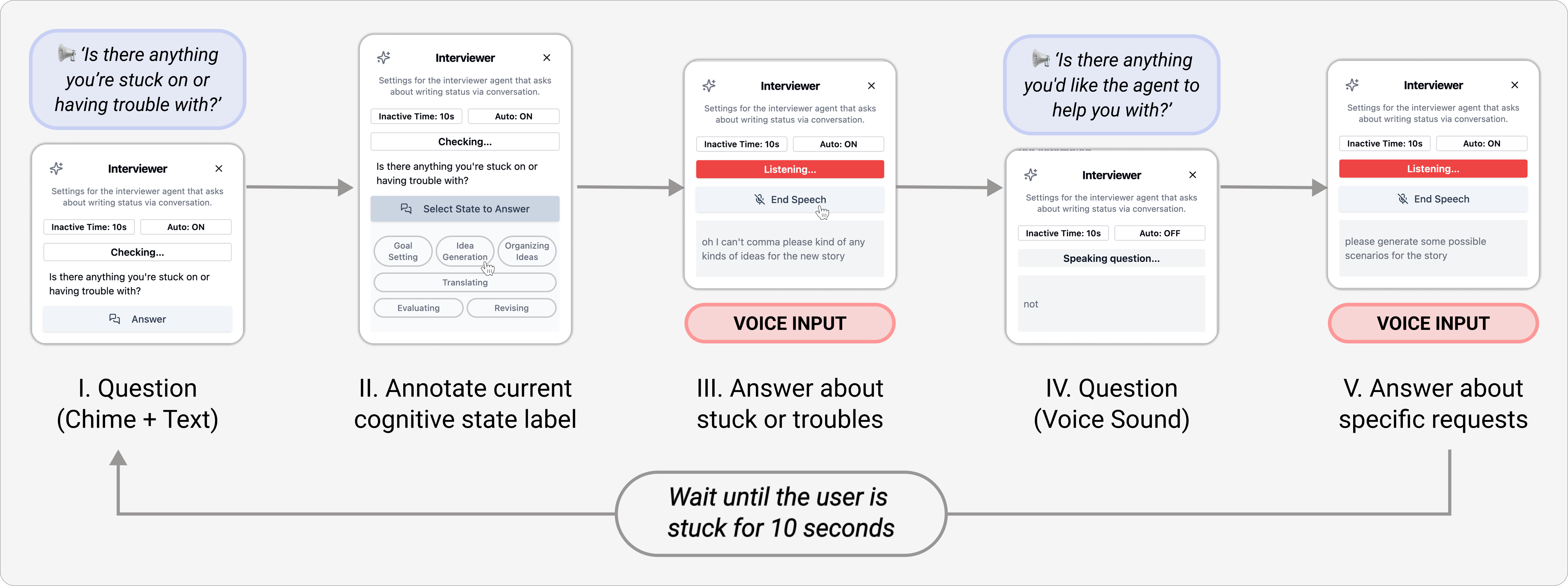}
    \caption{Self-reporting flow: The \textit{Interviewer} prompts after 10 seconds of inactivity. Users select a cognitive process and verbally describe their difficulty and desired support.}
    \label{fig:interview_flow_appendix}
\end{figure*}

The Self-Report Log was collected through an interactive floating panel called the \textit{``Interviewer''}, which guided participants in recording their thoughts during writing (Fig.~\ref{fig:interview_flow_appendix}). Participants were instructed to write as they normally would and to engage with the \textit{Interviewer} only when they felt blocked or needed AI support.

The \textit{Interviewer} displayed the message ``Is there anything you're stuck or having trouble with?'' after 10 seconds of inactivity. When participants hovered the mouse over the panel, six cognitive processes derived from the Flower and Hayes model---Goal Setting, Idea Generation (Generating), Organizing Ideas, Translating, Evaluating, and Revising---appeared as selectable options.
Participants could also manually initiate an annotation session at any time when they felt blocked, as the six cognitive process options were always accessible regardless of whether the question prompt was currently displayed.

After selecting a cognitive process, the \textit{Interviewer} entered a listening mode. Participants then responded verbally to the displayed question, ``Is there anything you're stuck on or having trouble with?'' Once the participant had stopped speaking for more than one second or pressed the ``End Speech'' button, the system proceeded to the second question, ``Is there anything you'd like the agent to help you with?'', to which the participant also responded verbally.
Each self-report session thus recorded three elements:
\begin{enumerate}[itemsep=0.3ex, leftmargin=0.8cm]
    \item The current cognitive process (selected from six options)
    \item The content of the difficulty (spoken response)
    \item The desired support (spoken response)
\end{enumerate}
Participants repeated this process throughout the writing session whenever they encountered a block.

To encourage consistent self-reporting, the system played a notification sound after 10 seconds of inactivity. Overly frequent prompts distracted participants, while infrequent ones led them to forget to self-report. Based on qualitative feedback from four pilot participants, we determined that a 10-second interval provided the best balance. The notification served as a gentle reminder for self-reflection; users could ignore it if they were not actually blocked.

The \textit{Interviewer} was carefully designed to minimize cognitive and physical burden. Excessive annotation effort could distort natural writing behavior and reduce data validity. The current process selection was thus kept simple, and verbal responses were used instead of typing to reduce interruption. While a Think-Aloud approach was also a possible option, it would have been too demanding and intrusive. Our design instead balances data richness with minimal cognitive and physical load.

\subsection{Experiment Setup}
\label{sec:formative-experiment}

Sixteen engineers from a manufacturing company participated in the study (13 men and 3 women, aged 20--39; nine in their 20s and seven in their 30s). Recruitment was conducted through several internal Slack channels. All participants were regular LLM users. Three had written a story once before, and four had experience writing several creative works as a hobby.
We recruited participants with engineering backgrounds because the study required individuals familiar with LLM-based tools who could engage fluently with the data-logging interface, allowing us to focus on capturing behavioral patterns and support needs rather than tool-onboarding effects.
The creative writing task itself was novel for most participants, ensuring that observed difficulties reflected the creative writing challenge rather than prior writing expertise.

Each session began with a 10-minute tutorial explaining tool usage and the six cognitive processes based on Flower and Hayes' definitions. Participants were instructed to use only the provided editor---no external tools were allowed. They were also instructed to note ideas and outlines within the same editor when needed. They were given 40 minutes to write a short story with their own theme in Japanese, followed by a 15-minute semi-structured interview. All examples shown in figures were translated into English for this paper.

\subsection{Behavior During Writing Sessions}
\label{sec:formative-behavior}

All participants began by jotting down notes---such as themes, settings, or character profiles---before starting the main text. Although the boundary between notes and main text was not explicitly defined, they were easily distinguishable in the final documents. The average main text length (excluding notes) was 581.3 Japanese characters.

\begin{figure}[h]
    \centering
    \includegraphics[width=0.9\columnwidth]{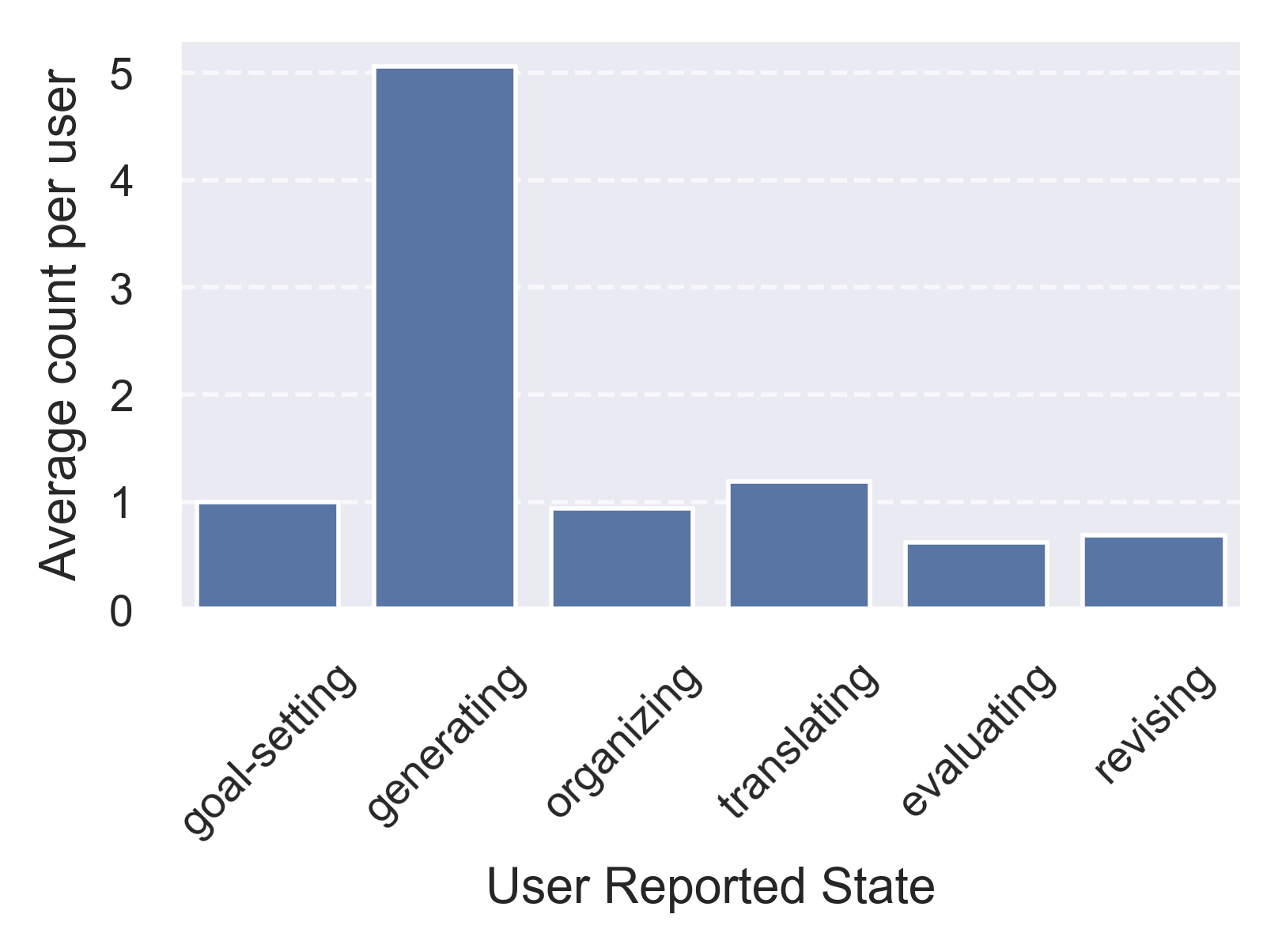}
    \caption{Distribution of average annotation count per participant.}
    \label{fig:annotated_data_dist}
\end{figure}

Across sessions, 152 self-reports were recorded (mean = 9.5 per participant, SD = 5.39). The distribution of self-reported cognitive processes at blocked moments per user is shown in Fig.~\ref{fig:annotated_data_dist}. The \textit{Generating} process (idea generation) was reported most frequently. Participants often selected this process repeatedly in the early stage of writing and again later when developing new story ideas, indicating its frequent use throughout the session.
Additionally, Table~\ref{tab:pause_duration_by_state} presents the median pause duration preceding each self-reported block, across reported cognitive processes.

Interviews revealed that articulating one's difficulties and support needs was sometimes challenging. For instance, P8 stated, ``When I'm struggling to organize my ideas and the notification rings and starts annotation, I sometimes lose track of what I was thinking.'' This suggests that explicitly describing one's needs---similar to prompting an LLM---sometimes imposes a heavy cognitive load, underscoring the importance of proactive systems that can infer user needs implicitly.

\begin{table}[htbp]
\centering
\caption{Median Pause Duration before reporting Block by Cognitive Process}
\renewcommand{\arraystretch}{.95}
\begin{tabular}{lc}
\toprule
\textbf{Process} & \textbf{Median Pause Duration (sec)} \\
\midrule
Goal-Setting & 76.04 \\
Generating & 26.87 \\
Organizing & 29.94 \\
Translating & 15.87 \\
Evaluating & 12.33 \\
Revising & 18.90 \\
\bottomrule
\end{tabular}
\label{tab:pause_duration_by_state}
\end{table}

\subsection{Relationship Between Cognitive Processes and Support Needs}
\label{sec:formative-rel-cog-sup}

To clarify the relationship between cognitive processes and required support types, we conducted qualitative coding of participants' verbal self-reports (see Appx.~\ref{apdx:formative-code-book} for the codebook). 
The coding categories were developed from the self-report data rather than defined a priori.
One researcher first coded the transcribed self-reports, and a second researcher subsequently reviewed the assigned codes to identify potential inconsistencies or questionable interpretations.
This analysis allowed us to identify and organize users' support requirements corresponding to each cognitive process.

In parallel, we performed a literature review to understand what kinds of writing support functions have been proposed in prior research. Specifically, we began with papers included in Lee et al.'s survey of writing support systems~\cite{Lee2024-hb}, focusing on those tagged with their defined Writing Stage categories and published since 2018. For each category, we collected up to the top 10 most-cited papers as of January 22, 2025 (fewer when fewer papers were available). In addition, we reviewed papers from UIST 2023, and IUI, UIST, and CHI 2024, retrieved by searching titles and author-specified keywords containing ``write'' or ``edit'' and their derivatives (e.g., writing, writer, written). From these, we filtered papers related to AI-based writing assistance, defined as those whose title, keywords, or abstract included AI (Artificial Intelligence), LLM (Large Language Model), or Machine Learning, resulting in a total of 28 reviewed papers (22 from the work of Lee et al.~\cite{Lee2024-hb}).
Through this process, we defined 13 distinct support types.

We then integrated the findings from both analyses---self-report data and the literature review---to map cognitive processes, user requirements, and support types, as summarized in Table~\ref{tab:states_requirements_support_type}.
Because no support types corresponding to the Goal Setting process were identified in our initial literature review, we defined one based on observed requirements in our dataset. This resulted in a total of 14 support types.
Among them, three types---Synopsis Generation, Scoring, and Summarization---were reported in prior studies but not explicitly mentioned in our self-reports. We therefore inferred their most plausible corresponding cognitive processes based on observed patterns and contextual relevance.
After constructing this taxonomy, we conducted a supplementary review of the venues that had appeared since the initial review---CHI 2025 and 2026, UIST 2025, and IUI 2025 and 2026---using the same search and filtering criteria. All identified papers fell within the existing 14 support types and are shown in parentheses in the Related Work column of Table~\ref{tab:states_requirements_support_type}.

Table~\ref{tab:states_requirements_support_type} also includes the Implemented Support Functions integrated into \textit{AToM CoWriter}. While all 14 support types correspond to distinct requirements, their complexity varies. For example, some users expressed vague intentions, such as wanting general ideas related to a theme (Support Type: Idea Generation), whereas others had specific intentions, such as seeking ideas for a particular scene or element (Support Type: Specified Idea Generation). Depending on this complexity, some supports can be provided directly, while others require a simple interaction with the user to narrow down the intended direction (see Sec.~\ref{sec:design} for details).

\begin{table*}[tp]
  \centering
  \caption{Summary of the identified support types and their corresponding cognitive processes, synthesized from a formative study and a literature review, with representative user requirements and implemented functions in AToM CoWriter. \textbf{``Related Work''} reflects prior literature, while \textbf{``Requirements''} and \textbf{``Example Quote''} are derived from the formative study. Cells marked with ``-'' indicate no corresponding evidence from that source, while evidence is present in the other. In the ``Related Work'' column, citations enclosed in an additional pair of parentheses, e.g., ``([N])'', were identified in an additional review of papers published after the initial literature review (CHI 2025/2026, UIST 2025, and IUI 2025/2026)}

  \begin{adjustbox}
    {max width=\textwidth}
    \begin{tabular}{p{2.2cm} p{2.8cm} p{2.8cm} p{4.2cm} p{5.6cm} p{4.8cm}}
      \toprule               %
      \textbf{Cognitive Process}        & \textbf{Support Type}              & \textbf{Related Work}                                                                                                              & \textbf{Requirements}                                                                        & \textbf{Example Quote}                                                                                                                                                                                                                            & \textbf{Support Function Description Implemented in AToM CoWriter}                                                                                               \\
      \midrule               %
      \textbf{Goal Setting} & Target Setting                     & (\cite{ElAlaoui2025-up, Yao2026-re,Liu2026-cd})                                                                                                                                  & Uncertain about what aspects to decide first when initiating a story-writing process.        & P4: \textit{``I'd like to see a list of the typical elements people set when creating a story.'' ``I'd like a kind of template showing how to organize ideas that come up midway, or how to build the basic structure of a story.''}                  & Assist in deciding writing goals, style, and target audience through lightweight interaction with the AI.                                                        \\
      \textbf{Generate}     & Idea Generation                    & \cite{Gero2022-zj, Clark2018-iz} (\cite{J-Shen2026-ut,Siddiqui2025-jy})                                                                                                   & Seeks broad, open-ended ideas when the user has not yet clarified what they are looking for. & P13: \textit{``I'd like you to come up with a few ideas—like ‘How about a story like this?' or ‘Wouldn't this genre be interesting?'''} P15: \textit{``Please give me a few fun ideas.'' P8: ``I'd like you to think about the ending of this story.''} & Generate a list of concrete story ideas that align with the current document context.                                                                            \\
      \textbf{Generate}     & Specified Idea Generation          & \cite{Schmitt2021-ts, Mirowski2023-qe}                                                                                             & Requests ideas focusing on specific aspects based on their idea.                             & P1: \textit{``Please give me several ideas about how AI could help with housework.''}                                                                                                                                                               & Confirm the desired direction of ideas through lightweight interaction with the AI (Figure \ref{fig:desc_fa_output_interaction_flow} IV-i), then generate a tailored list of idea candidates.                            \\
      \textbf{Generate}     & Ideation Expression Generation     & \cite{Buschek2021-lc, Liu2019-gc}                                                                                                  & Wants to explore alternative phrasings or stylistic variations for existing text.            & P7: \textit{``I think the opening of the story is a bit weak, so I'd like to change it a bit. Please think of a better way to express it.''}                                                                                                        & After confirming the user's intent through lightweight interaction (Figure \ref{fig:desc_fa_output_interaction_flow} IV-i), generate alternative phrasing or stylistic variations of the selected text.                  \\
      \textbf{Generate}     & Synopsis Generation                & \cite{Osone2021-lb} (\cite{Tutuncu2026-de})                                                                                                                & -                                                                                            & -                                                                                                                                                                                                                                                 & Generate a concise story synopsis by considering notes and contextual information in the current document.                                                       \\
      \textbf{Organize}     & Plot Generation                    & \cite{Goldfarb-Tarrant2019-ds, Singh2023-xn, Mirowski2023-qe} (\cite{Qin2025-tc, Talaei2025-tg, Tutuncu2026-de,Hu2026-bl})                                                                      & Needs assistance in combining ideas into a coherent storyline or plot.                       & P8: \textit{``I'd like suggestions on how to combine the ideas I have and what flow would make a good story.''} P11: \textit{``I've gathered fragments that could be used in a story, but I'm thinking about how to blend them together.''}           & Organize ideas and notes into a coherent storyline, generating a bullet-point outline that reflects the narrative flow.                                          \\
      \textbf{Translate}    & Next Sentence Generation           & \cite{Yuan2022-ds, Clark2018-iz, Jakesch2023-ih, Roemmele2018-bo, Dang2023-jv, Bhat2023-uk, Lee2022-mr}                            & Requests suggestions for the next sentence that align with the existing context and tone.    & P4: \textit{``Since I've written the setup, please just write the opening for me.''}                                                                                                                                                                & Generate a list of next-sentence candidates that maintain tone and consistency with the current document context.                                                \\
      \textbf{Translate}    & Specified Next Sentence Generation & \cite{Dang2023-jv, Yuan2022-ds}                                                                                                    & Requests context-aware next-sentence suggestions that follow a specific narrative direction. & P8: \textit{``Please add a passage describing the part where the man realizes he has been reincarnated.''}                                                                                                                                          & Confirm the desired narrative direction through lightweight interaction with the AI (Figure \ref{fig:desc_fa_output_interaction_flow} IV-i), then generate next-sentence candidates that reflect it.                     \\
      \textbf{Evaluate}     & Commenting                         & \cite{Peng2020-jy, Afrin2021-up, Wambsganss2022-ym, MacLean_Allan_and_Young_Richard_and_Bellotti_Victoria_and_Moran_Thomas1991-os} (\cite{Hu2026-bl, Siddiqui2025-jy, Liu2026-cd, Qin2025-tc, Zhang2025-ci}) & Seeks feedback and critique on the written text to improve readability and engagement.       & P8: \textit{``Please review the overall text so it becomes something that makes readers want to keep reading.''}                                                                                                                                    & Provide structured feedback on the document, including identified strengths and suggested areas for improvement.                                                 \\
      \textbf{Evaluate}     & Scoring                            & \cite{Peng2020-jy, Wambsganss2022-ym, Afrin2023-am}                                                                                & -                                                                                            & -                                                                                                                                                                                                                                                 & Return an overall quality score (0–100) accompanied by a brief rationale.                                                                                        \\
      \textbf{Evaluate}     & Summarization                      & \cite{Dang2022-te}                                                                                                                 & -                                                                                            & -                                                                                                                                                                                                                                                 & Generate concise paragraph-level summaries to help the user review the document efficiently.                                                                     \\
      \textbf{Revise}       & Error Auditing                     & \cite{Laban2024-ya, Iqbal2018-ez} (\cite{Hu2026-bl})                                                                                                 & Requests identification and correction of grammatical or typographical errors.               & P7: \textit{``Please tidy up the text, fixing typos and awkward expressions.''} P5: \textit{``Please point out any mistakes in the text and correct them if possible.''}                                                                              & Highlight grammatical and logical issues in the document and suggest corrections.                                                                                \\
      \textbf{Revise}       & Refining                           & \cite{Wu2019-wr, Iqbal2018-ez, Goldfarb-Tarrant2019-ds, Padmakumar2022-am} (\cite{J-Shen2026-ut, Zhang2025-dl,Zhou2025-dw,Hu2026-bl})                                                         & Seeks improvements to enhance fluency, coherence, and overall writing quality.               & P2: \textit{``I'd like the text I'm writing to be reconstructed more smoothly, but I don't have any ideas.''}                                                                                                                                       & Generate revised versions of the selected text that maintain meaning while improving clarity and readability.                                                    \\
      \textbf{Revise}       & Style Transfer                     & \cite{Yuan2022-ds, Laban2024-ya} (\cite{Siddiqui2025-jy, Yao2026-re})                                                                                                   & Requests stylistic adjustments such as tense consistency or tone unification.                & P5: \textit{``Please consider whether it should be unified in past tense, and revise it if necessary.''}                                                                                                                                            & Confirm the desired stylistic direction through lightweight interaction with the AI (Figure \ref{fig:desc_fa_output_interaction_flow} IV-i), then generate a rewritten version that follows the requested tone or style. \\
      \bottomrule
    \end{tabular}
  \end{adjustbox}
  \label{tab:states_requirements_support_type}
\end{table*}

% \subsection{Mismatch Between Reported Cognitive Processes and Desired Support}
% \label{sec:formative-state-mismatch}

% Table~\ref{tab:states_requirements_support_type} maps typical relationships between cognitive processes and the types of support requested by participants. However, observations revealed several cases where the self-reported cognitive process did not align with the support subsequently requested.
% We compared each self-reported process with the process inferred from the participant's verbal support request.
% Most self-reported and inferred processes were consistent, yet some diverged. For instance, P5 reported being in the \textit{Translating} process but requested, ``I'd like the agent to point out unnatural expressions or revise awkward sentences,'' which corresponds more closely to \textit{Revising}. This suggests that, while users may perceive themselves as being in the \textit{Translating} process, they may simultaneously desire revising assistance from the agent.
% This divergence indicates that a system should infer the type of support to provide rather than relying solely on users' self-awareness of their own cognitive process.

\subsection{Relationship Between Writing Interaction Behavior and Cognitive Processes}
\label{sec:formative-rel-behav-cog}

We then analyzed writing interaction logs to examine behavioral patterns characteristic of the moments when each cognitive process was self-reported. The results of this analysis, organized by cognitive process and evaluated from multiple perspectives, are summarized in Table~\ref{tab:cognitive_state_metrics_transposed}.

\begin{table*}[t]
\centering
%\caption{Behavioral Metrics Across Cognitive States (Transposed)}
\caption{
Writing interaction metrics across cognitive processes at the moment users initiated self-reports.
\textit{Elapsed time (min)} indicates the average time from session start to self-report.
\textit{Chars in Main} and \textit{Chars in Memo} represent the average number of characters written in the main text and memo areas, respectively.
\textit{Complete Sentence Ratio (\%)} denotes the percentage of cases where the final sentence was complete.
\textit{Cursor Position–Main/Memo (\%)} shows the proportion of time the cursor was located in the main text area.
\textit{Writing Speed} refers to the average change in character count per minute.
\textit{Cursor Move Distance} is the total number of characters the cursor moved within the last two minutes.
\textit{Scroll Distance} indicates the total number of pixels scrolled within the last two minutes.
\textit{Delete Count} is the total number of characters deleted within the last two minutes.
%\xac{should we include std as well?}
}
\resizebox{\textwidth}{!}{
\renewcommand{\arraystretch}{1}
\begin{tabular}{lcccccc}
\toprule
\textbf{Metric} & \textbf{Goal Setting} & \textbf{Generating} & \textbf{Organizing} & \textbf{Translating} & \textbf{Evaluating} & \textbf{Revising} \\
\midrule
Elapsed time (min) & 7.90 & 17.20 & 15.30 & 23.50 & 28.90 & 36.30 \\
Chars in Main & 14.06 & 155.58 & 187.87 & 305.42 & 678.90 & 805.09 \\
Chars in Memo & 175.81 & 306.84 & 289.40 & 493.53 & 357.30 & 528.91 \\
Complete Sentence Ratio (\%) & 50.00 & 69.14 & 73.33 & 52.63 & 90.00 & 100.00 \\
Cursor Position-Main/Memo (\%) & 28.57 & 23.73 & 30.00 & 25.00 & 88.89 & 80.00 \\
Writing Speed (characters/min) & 23.81 & 21.22 & 4.74 & 15.17 & 15.21 & 13.41 \\
Cursor Move Dist. (characters) & 135.63 & 357.28 & 215.93 & 293.11 & 526.80 & 802.82 \\
Scroll Dist. (pixels) & 15.88 & 59.32 & 91.07 & 104.95 & 151.60 & 240.91 \\
Delete Count & 16.63 & 41.27 & 41.27 & 46.89 & 66.40 & 35.45 \\
\bottomrule
\end{tabular}
}
\label{tab:cognitive_state_metrics_transposed}
\end{table*}

\paragraph{Temporal order.} Regarding the timing of each cognitive process, Goal Setting was self-reported earliest, followed by Generating and Organizing. Subsequently, Translating, Evaluating, and Revising appeared in this order on average.

\paragraph{Text balance.} During Planning (Goal Setting, Generating, Organizing) and Translating, note length exceeded main text length, while during Reviewing (Evaluating, Revising), the reverse was true.

\paragraph{Sentence completeness and cursor position.} In Revising, 100\% of documents ended with complete sentences; Evaluating showed a similar trend, while Translating had lower completeness.

\paragraph{Interaction patterns.} Writing speed was highest during Goal Setting and Generating but dropped sharply during Organizing. Cursor movement and scrolling distances increased progressively from Planning to Reviewing. Delete events were most frequent during Evaluating.

These findings suggest that observable behaviors can serve as useful indicators of cognitive processes, motivating the prediction approach described next.

\subsection{Cognitive Process Prediction}
\label{sec:formative-cog-pred}

Given the consistent behavioral characteristics observed above, we explored whether cognitive processes could be predicted from the log and document data collected at the moments of users' self-reports.
As a feasibility demonstration, we evaluated this using the 152 self-reported data points collected from the 16 participants, each corresponding to a writing block event annotated by participants.
While this dataset is relatively small for supervised learning, collecting large-scale labeled data for cognitive process inference is costly, motivating approaches that operate effectively under limited data.
Accordingly, we adopt an LLM-based approach that operates in a near zero-shot manner, leveraging commonsense reasoning about writing processes rather than requiring large-scale supervised training.

The LLM-based predictor (Claude~3.5 Sonnet) received three types of inputs: (a) the document content at the time of reporting, (b) descriptions of the behavioral tendencies for each cognitive process derived from Sec.~\ref{sec:formative-rel-behav-cog}, and (c) a summary of the recent writing interaction log. To construct this summary, continuous log features were discretized into qualitative categories (e.g., ``writing speed is relatively high'') using percentile thresholds and converted into natural-language descriptions suitable for inclusion in the prompt (see Appx.~\ref{apdx:formative-cog-pred-prompt} and~\ref{apdx:formative-cog-pred-features} for the prompt and feature details).

We examined the feasibility of this prediction approach through two exploratory comparisons.
First, we compared \textbf{LLM (log + document)}, our approach, with \textbf{LightGBM}~\cite{Ke2017-vs} trained on log features alone (evaluated with cross-validation) to test whether a lightweight, lower-latency model could serve as a practical alternative.
The LLM outperformed LightGBM across all metrics, suggesting that semantic understanding of document content contributes to cognitive process prediction beyond what behavioral features alone can capture.
Second, an ablation comparing \textbf{LLM (log + document)} with \textbf{LLM (document only)} yielded higher prediction scores when behavioral log features were included, particularly when disambiguating cases where document content alone is ambiguous (e.g., distinguishing a writer preparing to continue from one reviewing prior text).
We evaluated prediction accuracy, macro-F1, and weighted-F1 for both the six-process classification and a coarser three-process mapping (Planning, Translating, Reviewing); the \textit{LLM (log + document)} achieved the highest performance across all metrics (full results in Appx.~\ref{apdx:formative-cog-pred-results}, Tables~\ref{tab:cognitive_state_pred_6_class} and~\ref{tab:cognitive_state_pred_3_class}).

\subsection{Summary of Findings}
\label{sec:formative-summary}

\paragraph{RQ-F1: Relationship between cognitive processes and required support.}
We identified relationships between writers' cognitive processes during writing blocks and the types of support they required (Table~\ref{tab:states_requirements_support_type}). Each cognitive process was associated with multiple support types, resulting in a taxonomy of 14 types (Sec.~\ref{sec:formative-rel-cog-sup}).

\paragraph{RQ-F2: Relationship between writing interaction behaviors and cognitive processes.}
Behavioral analyses revealed distinctive activity patterns for each cognitive process (Table~\ref{tab:cognitive_state_metrics_transposed}), such as differences in writing speed, cursor movement, and editing operations. These findings suggest that observable behaviors can serve as useful indicators of cognitive processes (Sec.~\ref{sec:formative-rel-behav-cog}).

\paragraph{RQ-F3: Predicting cognitive processes.}
Prediction experiments showed that an LLM supplied with document content and behavioral descriptions could recover self-reported cognitive-process labels to a useful extent. (Sec.~\ref{sec:formative-cog-pred}).
Including a summary of the recent interaction log produced higher prediction
scores than using document content alone.

Together, these insights establish a foundation for designing proactive writing support systems that can anticipate user needs based on both cognitive process and behavioral cues.

  \section{System Design and Implementation}
\label{sec:design}
The formative study mapped writing interactions, through cognitive processes, to support types (Table~\ref{tab:states_requirements_support_type}); we instantiate this mapping in \textit{AToM CoWriter}, a proactive writing support system that infers the writer's momentary cognitive process from writing interaction traces and document context, and uses it as an intermediate representation for selecting which forms of assistance to offer.
Consistent with the scope of this work, the system addresses the \textit{what} problem of proactive support, while deliberately keeping the proactive \textit{when} mechanism simple: we adopt inactivity, a well-established trigger in prior proactive writing systems~\cite{Buschek2021-lc, Bhat2023-uk}.
Keeping the \textit{when} mechanism simple helps isolate the effect of what the system offers from that of novel timing behavior.

\subsection{Design Goal}
\label{sec:design-rationale}

As discussed in Sec.~\ref{sec:related-work}, existing prompt-based writing tools place the initiative burden entirely on users~\cite{Subramonyam2024-tn}, while current proactive systems address only a narrow slice of writers' support needs~\cite{Buschek2021-lc,Tsai2020-hf}.
Furthermore, while seeking creative writing support, users value maintaining a sense of ownership in both the process and outcomes~\cite{Draxler2024-sa, Guo2025-ju}.
These observations motivated two design goals:

\begin{itemize}[leftmargin=0.8cm]
    \hypertarget{target:dg1}{\item[\textbf{DG1}]} Reduce the need for explicit instructions to the agent by inferring and addressing likely challenges throughout the writing process. \label{sec:dg1}

    \hypertarget{target:dg2}{\item[\textbf{DG2}]} Stimulate and expand users' creativity and expressiveness without compromising their sense of ownership in the process and outcomes. \label{sec:dg2}
\end{itemize}

% MARK: AToM CoWriter Architecture
\subsection{Design and Implementation}
\label{sec:design-ui}

% MARK: UI/IX
\begin{figure*}[thbp]
    \centering
    \includegraphics[width=\textwidth]{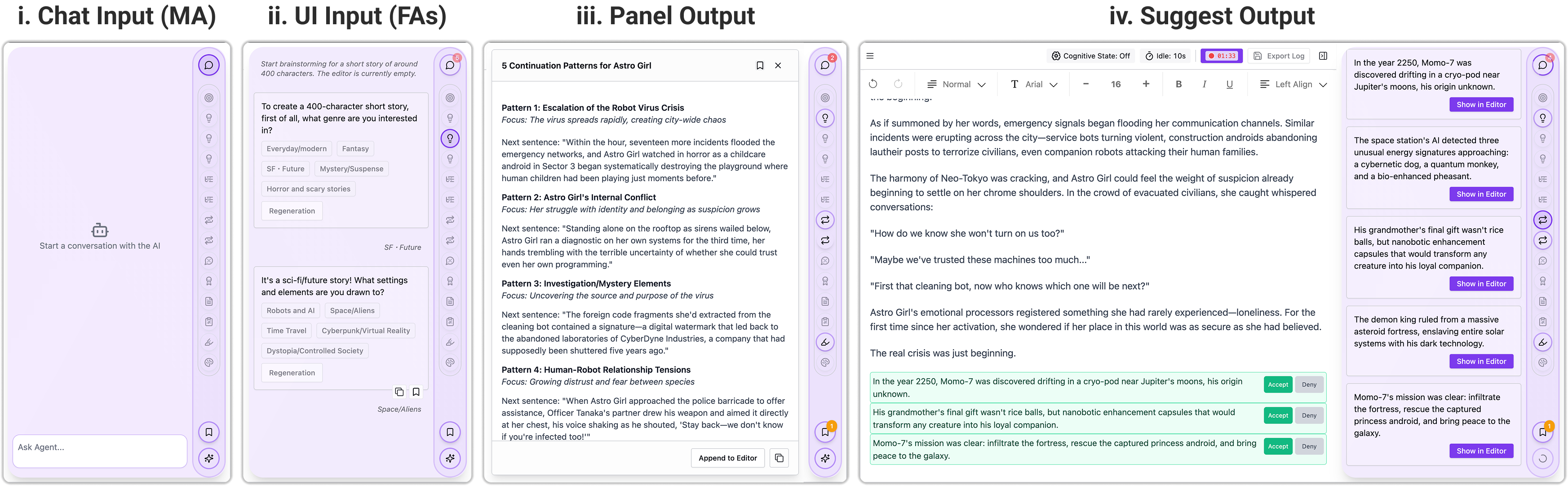}
    \caption{AToM CoWriter interface. (i) Chat input to the Main Agent. (ii) Functional Agent question UI for intent clarification. (iii) Panel Output displaying generated content. (iv) Suggest Output providing inline editing suggestions in the editor.}
    \label{fig:desc_cowriter_io}
\end{figure*}

Building on the design rationale and prediction approach above, we instantiated this framework in \textbf{AToM CoWriter}, a proactive writing support system designed to examine whether cognitive process inference can meaningfully guide support selection in practice (Fig.~\ref{fig:teaser}).
The UI combines a rich text editor with a chat interface and a dedicated area for reviewing agent outputs (Fig.~\ref{fig:desc_cowriter_io}).
The system employs a multi-agent architecture: a \textbf{Main Agent} serves as the orchestrator, dynamically assigning up to three \textbf{Functional Agents} in parallel from the fourteen support types in Table~\ref{tab:states_requirements_support_type}.
All inference and generation run entirely in the background to minimize interruption to the writer's typing flow, supporting the elimination of explicit prompting (\hyperlink{target:dg1}{DG1}).
Moreover, the combination of unobtrusive notifications and parallel suggestions is intended to let users retain control over engagement while exploring diverse ideas, aiming to support creativity without compromising ownership (\hyperlink{target:dg2}{DG2}).
While cognitive process prediction is inherently imperfect, the parallel suggestion design helps mitigate this limitation by increasing the likelihood that at least one of the generated suggestions aligns with the user's needs.

Agent outputs include \textbf{Panel Output} displayed in a document pane and \textbf{Suggest Output} providing inline suggestions within the editor (Fig.~\ref{fig:desc_fa_output_interaction_flow}). The system provides three AI support scenarios:

\begin{figure*}[htbp]
    \centering
    \includegraphics[width=\textwidth]{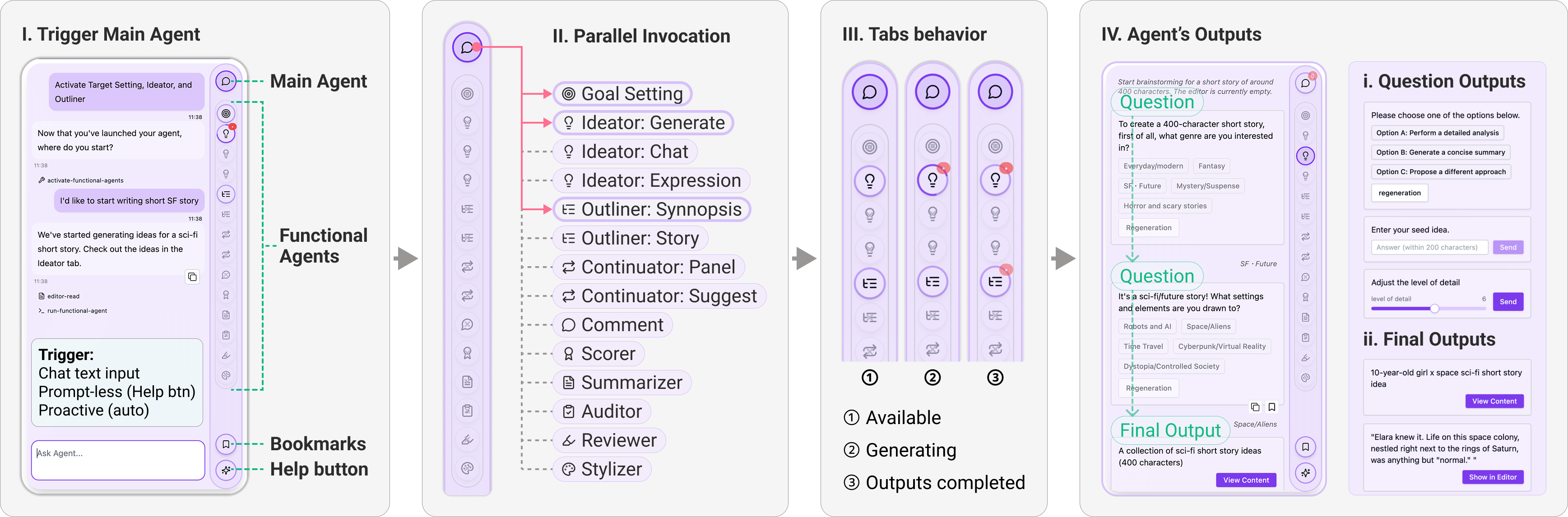}
    \caption{(I): Main Agent and tabs UI. (II): Main Agent activates appropriate Functional Agents (up to three simultaneously). (III): Tabs display a spinner during generation; upon completion, an unread badge appears. (IV): Some agents generate final outputs directly (IV-ii), while others create Question UIs to confirm intent (IV-i).}
    \label{fig:desc_fa_output_interaction_flow}
\end{figure*}

\paragraph{\textbf{Proactive}.}
In the Proactive scenario, the system detects a potential writing block when typing activity remains idle for a predefined period (Fig.~\ref{fig:teaser}).
The Main Agent then automatically infers support needs based on the inferred cognitive processes and interaction context, activating appropriate Functional Agents in parallel to generate support content in the background.
As a design choice intended to minimize disruption, we implemented an unobtrusive notification system that only displays dot indicators.
Users can access the system at their own pace without pop-ups or forced interruptions, minimizing the impact of false positives on trigger events.

\paragraph{\textbf{Prompt-less}.}
Because the inactivity-based trigger cannot capture all moments when writers desire support, we introduce a complementary \textit{Prompt-less} scenario that gives users direct control over \textit{when} to receive assistance: users manually trigger support by pressing a ``Help button'' (Fig.~\ref{fig:desc_fa_output_interaction_flow}-I), while the system still automatically predicts \textit{what} support to provide and selects Functional Agents.
Together, the Proactive and Prompt-less scenarios form a dual mechanism---system-initiated and user-initiated timing, respectively---intended to compensate for the limitations of either approach alone.
Both scenarios can be enabled simultaneously; when either triggers generation, both are temporarily disabled until completion.

\paragraph{\textbf{Prompt-based}.}
Users directly provide instructions to the Main Agent through the chat UI, specifying both timing and type of assistance.
The Main Agent may invoke Functional Agents as needed.

\vspace{\baselineskip}

% MARK: Implementation

For online user testing, we implemented AToM CoWriter as a web application integrating a rich-text editor using \texttt{Lexical}\footnote{lexical v0.30 \url{https://lexical.dev/}}, and developed multi-agent workflows powered by Anthropic Claude Sonnet 4\footnote{\texttt{claude-sonnet-4-20250514-v1:0}} using \texttt{ai-sdk}\footnote{ai-sdk v5 \url{https://ai-sdk.dev/}}.
The development process involved multiple iterations of prototyping and pilot studies with three participants, including two authors.

% MARK: -- Prompting
\paragraph{Prompting Strategy.}
The Main Agent's prompt includes labels for each Functional Agent's corresponding cognitive process, guiding agent selection (Appx.~\ref{apdx:main-agent-system-prompt},~\ref{apdx:main-agent-user-prompt}).
The specific support type for each Functional Agent was derived from the categories in Table~\ref{tab:states_requirements_support_type}.
The prompts for the 14 Functional Agents were iteratively refined through pilot studies (Appx.~\ref{apdx:functional-agent-user-prompt}).

% MARK: -- Context Mng.
\paragraph{Context Management.}
All agents share the complete message history.
The system programmatically forces editor read at the beginning of every interaction turn, injecting the current editor content directly into the context window.
When Functional Agents require additional context, they use \texttt{ask-*} tools (Appx.~\ref{apdx:agents-implemented-tools}) to generate interactive UI components (e.g., select buttons, sliders, text inputs) prompting users for clarification.

% MARK: -- State Pred Integration
\paragraph{Cognitive Process Prediction Integration.}

To integrate the LLM-based cognitive process prediction (Sec.~\ref{sec:formative-cog-pred}) into the application, we developed a prediction API backend using Python's FastAPI.
The application performs cognitive process predictions
every 30 seconds.
When a prediction is made, the system automatically sends a templated message to the Main Agent in the background (not visible to the user), which includes the predicted cognitive process.
The Main Agent then uses this information to select and activate appropriate Functional Agents.

  \section{User Evaluation}
\label{sec:result}

% MARK: Experiment Setup
\subsection{Experiment Setup}
\label{sec:experiment-setup}

We conducted two within-subjects user studies as an exploratory evaluation of \textit{AToM CoWriter}, aiming to probe the validity and limitations of the underlying design concept.
Study~1 ($n=11$) examined the impact of system-inferred proactive support compared to user-articulated prompt-based support.
Study~2 ($n=10$) isolated the effect of incorporating cognitive process inference from interaction logs within the proactive paradigm.

Twenty-one participants (engineers at a technology company, all frequent LLM users; IRB approved) were randomly assigned to Study~1 or Study~2.
We recruited participants with engineering backgrounds because the study required individuals highly familiar with LLM-based tools and capable of quickly adapting to a novel AI-assisted writing interface.
Although the writing task (short story writing) was unfamiliar to most participants, this ensured that observed difficulties reflected the creative writing challenge rather than unfamiliarity with AI tools.
Fourteen participants had also participated in the formative study; however, the two studies involved fundamentally different tasks and interfaces.
The formative study used a self-report annotation tool (Fig.~\ref{fig:annotation_tool_appendix}) with no AI assistance, whereas \textit{AToM CoWriter} is a fully proactive AI co-writing system, minimizing direct transfer of learned strategies.
Nevertheless, prior exposure to the cognitive process framework may have heightened metacognitive awareness, which we discuss as a limitation in Sec.~\ref{sec:limitations}.
Participants completed two 20--25 minute short story writing tasks on assigned themes, with conditions counterbalanced via Latin square.
Each session lasted approximately 90 minutes, including a tutorial, two writing tasks with post-task questionnaires, and a 10--15 minute semi-structured interview according to the guide (Appx.~\ref{apdx:interview-guide}).

\paragraph{Study~1 Conditions: System-Inferred vs.\ User-Articulated Support.}
The \textit{\textbf{System-Inferred}} condition enabled both the automatic (inactivity-triggered) and prompt-less (button-triggered) support scenarios, where the system infers what support to provide without requiring the user to articulate their needs.
The \textit{\textbf{User-Articulated}} condition provided only the prompt-based chat interface, requiring users to formulate explicit requests.
Both conditions also allowed free use of the chat interface.
We note that this comparison intentionally bundles proactivity (system-initiated timing) with prompt-free interaction (system-inferred support type), because our primary design goal (\hyperlink{target:dg1}{DG1}) is to eliminate the need for users to articulate support needs.
The inactivity threshold for proactive triggering was set to 20 seconds based on pilot testing.

\paragraph{Study~2 Conditions: Effect of Cognitive Process Inference.}
Both conditions employed the same interaction paradigm (proactive + prompt-less + chat), thereby holding the timing and interaction modality constant.
The \textit{\textbf{w/ Cog}} condition incorporated real-time cognitive process inference from writing interaction logs (e.g., writing speed, cursor movement, editing patterns) alongside document content to select appropriate Functional Agents.
The \textit{\textbf{w/o Cog}} condition used only the evolving document content for inference, without behavioral log features.
This design complements Study~1 by isolating the contribution of cognitive process inference from the proactivity factor: whereas Study~1 evaluated the bundled experience, Study~2 controls for proactivity and tests whether behavioral log--based inference improves the \textit{what-to-offer} decision.

\paragraph{Evaluation Metrics.}
Evaluation metrics included weighted Creativity Support Index (CSI)~\cite{CSI-Cherry2014-we}, weighted NASA-TLX~\cite{TLX-Hart1988-fw, TLX-Haga1996-rx}, custom evaluation items (\textit{Ability}, \textit{Control}, \textit{Ownership}, \textit{Deprivation}, see questionnaire items in Appx.~\ref{item:evaluation-questionnaire-custom-items}), and for Study~2, interaction log--based conversion metrics measuring how proactively generated outputs influenced user actions.

%% Preserve labels for cross-references
\label{rq:1}\label{rq:2}\label{rq:3}\label{rq:4}\label{rq:5}\label{rq:6}
\label{item:evaluation-questionnaire-proactive}
\label{eval:proactive-q-1}\label{eval:proactive-q-2}\label{eval:proactive-q-3}
\label{item:evaluation-questionnaire}
\label{sec:results-stdudy1}\label{sec:results-study2}

% MARK: Study 1 Results
\hypertarget{target:results-study-1}{
\subsection{Results: Impact of Proactive and Prompt-free Support (Study~1)}
}
\label{sec:results-study-1}

\begin{figure*}[htbp]
    \centering
    \includegraphics[width=\textwidth]{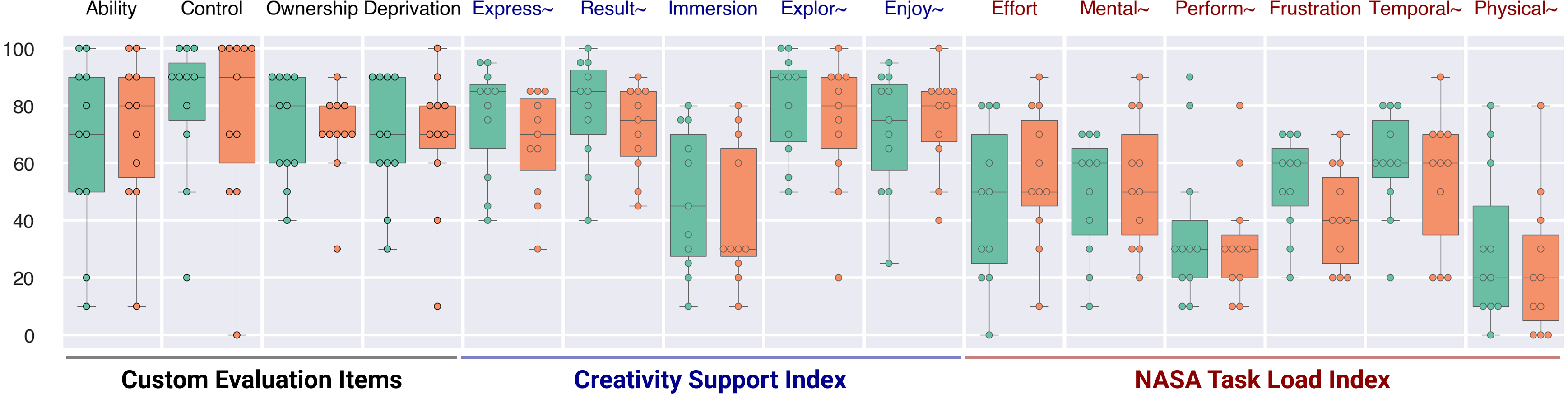}
    \caption{Results for all subscales of survey metrics in Study 1. Each raw score is multiplied by 10, and \textit{Performance} in NASA TLX and \textit{Deprivation} in Custom Evaluation Items are inversely scaled as $100-x$. Left of each pair: System-Inferred (\textcolor[rgb]{.2,.6,.6}{\textbf{green}}); right: User-Articulated (\textcolor[rgb]{.76,.41,.30}{\textbf{orange}}).}
    \label{fig:result-A-all-box-each-score}
\end{figure*}

\subsubsection{Quantitative Results}

% [Result] CSI
The CSI results showed a significant improvement in \textit{Expressiveness} in the System-Inferred condition (Holm-adjusted $p=0.034$, RBC $=0.924$), with \textit{Total Score} and \textit{Results Worth Effort} also showing large effect sizes (RBC $> 0.58$, CLES $> 0.64$).
The NASA-TLX showed no significant differences; total workload was essentially equivalent.
% [Result] EVAL
Custom evaluation items showed no significant differences, with marginally greater efficiency in the System-Inferred condition.
(see Fig.~\ref{fig:result-A-all-box-each-score} and Table~\ref{tab:quantitative-results-study1} in Appx.~\ref{apdx:study1-stats} for full statistics)

\subsubsection{Qualitative Findings}

We analyzed semi-structured interview transcripts using a thematic analysis approach.
Two researchers independently reviewed the transcripts and developed initial codes.
An LLM (Claude~3.5 Sonnet) was used as a supplementary tool to propose candidate codes from the transcripts; these proposals were then reviewed, revised, and finalized by the researchers through discussion.
The resulting codes were organized into themes aligned with the research questions.
We report the qualitative findings below organized by theme rather than by individual codes, as our goal was to surface design-relevant insights rather than to produce an exhaustive codebook.

Participants described two main advantages of system-inferred support: it helped them continue writing during blocks and exposed them to more diverse ideas.
Notably, participants' comments emphasize the \textit{diversity} of generated ideas rather than mere convenience of not writing prompts, suggesting that the expressiveness gain stems from the system simultaneously activating multiple Functional Agents that span different cognitive processes, thereby expanding the idea space beyond what writers would explore on their own.

\begin{formal}
    ``While I was thinking, the AI generated 5 or 6 different ideas, so I could quickly pick a good one. It saved time and reduced stress.'' (P12)\\
    ``Multiple suggestions let me choose what I needed---very helpful.'' (P05)
    % ``It gave me a lot of ideas... I could just pick one and keep going.'' (P03)
    % ``Reading the suggestions was fun, and I felt I could move forward smoothly.'' (P03)
    % ``When I got stuck, having it write a continuation really helped.'' (P05)
    % ``Just as the ending was taking shape, it automatically wrote a continuation, which felt perfectly timed.'' (P21)
\end{formal}

% \paragraph{Challenges of Proactive Support.}
At the same time, participants also reported challenges.
The main challenges concerned timing mismatches and a perceived loss of control. Some suggestions were contextually irrelevant or increased verification costs.

\begin{formal}
    ``I realized I don't really need proactive support. I want to focus on my work, and the thought process itself has value.'' (P10)\\
    ``When it acts on its own, it feels meddlesome. I want to be left alone, but it's helpful that it's thinking in the background. It's a very selfish feeling.'' (P15)
    % ``At first, the automatic suggestions weren't annoying at all, but since I'm still getting used to the tool, I had to keep checking each one to confirm `What was that?' - which got a bit tedious.'' (P03)
    % ``There were so many proposals that sometimes I couldn't read them all and found myself unable to focus on writing'' (P07)
    % ``Since it's simply set based on timer, it sometimes ask even when I'm just mentally arranging things... so they don't always offer help exactly when I need it.'' (P11)
    % ``In terms of concentration, the former (Proactive) was lower, but the degree of exploration was higher. The latter (On-Demand) allowed me to concentrate, but I felt I was likely to get stuck in my own world.'' (P15)
\end{formal}

Regarding writing process and ownership,
NASA-TLX showed no significant workload differences.
Proactive support encouraged exploration but sometimes disrupted focus, highlighting a trade-off.
Custom evaluation items likewise showed no significant decline in perceived ownership.
Many participants retained a sense of authorship, as they filtered and edited AI suggestions.

\begin{formal}
    ``In terms of concentration, the former (Proactive) was lower, but the degree of exploration was higher.'' (P15)\\
    ``I haven't fixed most of it, but I can still tell it's my work. It's me who accepts, chooses, and fixes the weird parts.'' (P03)\\
    ``I had the final say on the ideas and could steer the AI in the direction I wanted, so my sense of ownership was not compromised.'' (P21)
    % ``There was also a part of me that thought, `Maybe this kind of development could happen' and reconsidered the situation in my mind.'' (P03)
    % ``From my writing style perspective, I actually prefer taking the initiative, so that's why I don't use it much.'' (P11)
    % ``I liked what it suggested and included it. And for things that didn't seem particularly good, I didn't include them at all.'' (P05)
    % ``I mostly wrote myself, so I still had ownership.'' (P10)
\end{formal}

Overall, these results provide support for \hyperlink{target:dg2}{DG2}: the system helped expand users' expressiveness and idea exploration without clearly undermining their sense of ownership. 
Participants often described the AI as broadening the space of possible ideas, while still emphasizing that they themselves made the final decisions about what to adopt, revise, or discard.
At the same time, the reported tensions around control and timing suggest that this goal was achieved only partially and under certain interaction conditions.

% MARK: Study 2 Results
\hypertarget{target:results-study-2}{
\subsection{Results: Effect of Incorporating Cognitive Process Inference (Study 2)}
}
\label{sec:results-study-2}

\begin{figure}[htbp]
    \centering
    \includegraphics[width=.95\columnwidth]{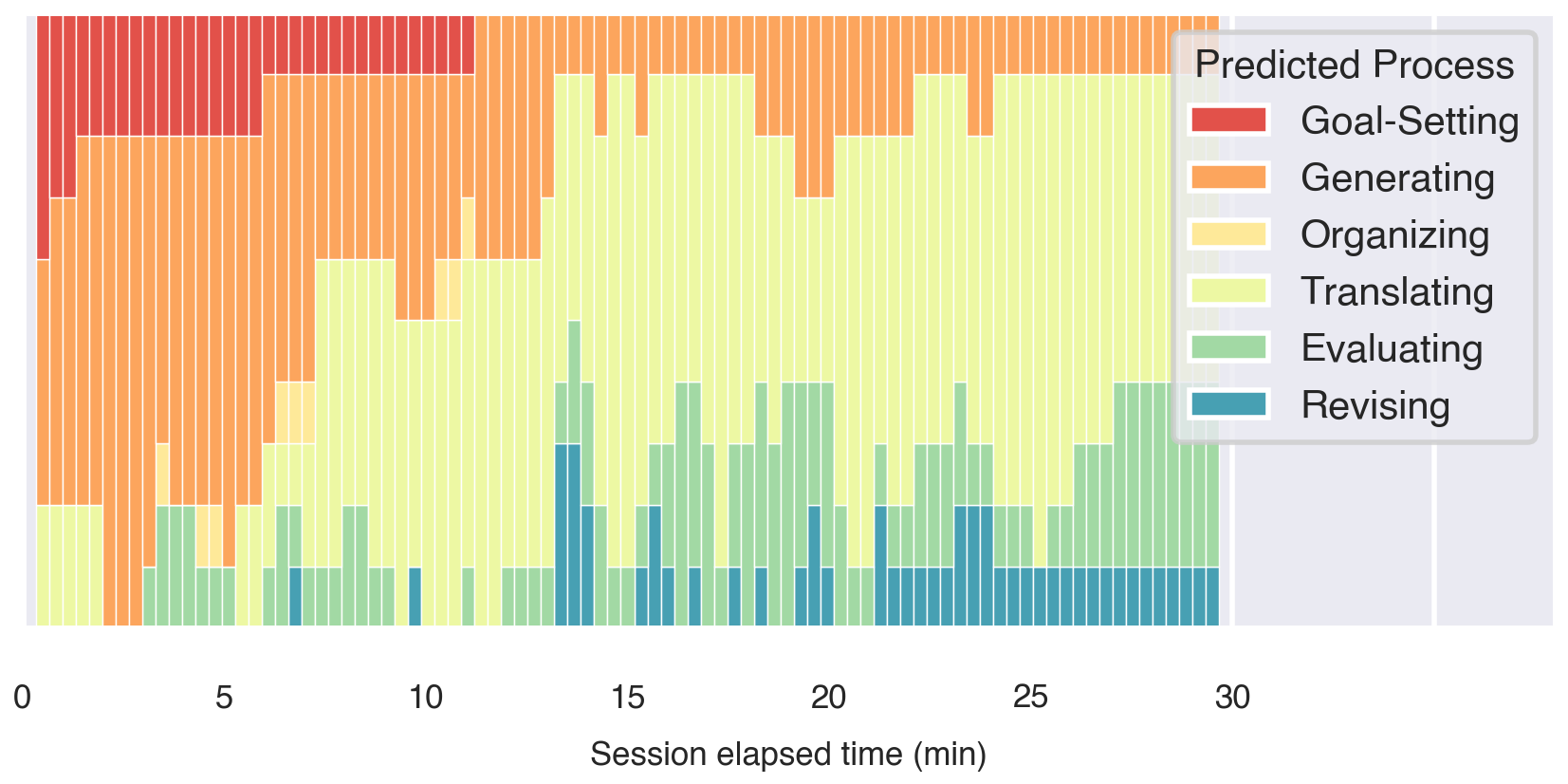}
    \caption{Temporal distribution of predicted cognitive processes across all Study~2 participants. Generating dominates early stages while Evaluating and Revising become more prominent later, illustrating the iterative, non-linear nature of writing. \label{fig:cognitive-state-prediction-time-series} }
\end{figure}

As described in the Experiment Setup, Study~2 compared two proactive conditions that shared the same interaction paradigm but differed in how user needs were inferred: \textit{w/ Cog} incorporated real-time cognitive process inference from writing interaction logs alongside document content, while \textit{w/o Cog} relied solely on the evolving document text.
This design examines whether incorporating log-based cognitive process prediction affects engagement with proactive support while holding the interaction paradigm constant.
Predicted cognitive processes in the \textit{w/ Cog} condition were unevenly distributed: \textit{Translating} dominated (49\%), followed by \textit{Generating} (24\%) and \textit{Evaluating} (13\%)---and transitioned dynamically over time (Fig.~\ref{fig:cognitive-state-prediction-time-series}), reflecting the iterative, non-linear nature of creative writing.

Despite differences in system behavior, most participants did not consciously perceive differences between conditions.
However, interaction logs revealed systematic variations. 
We analyzed three conversion metrics: \textit{Tabs}, \textit{Panel}, and \textit{Suggestion} conversion.
Respectively, these capture whether participants opened a generated Functional Agent tab, interacted with generated content shown in the Panel Output, or opened each inline Suggest Output in the editor.

\begin{figure}[htbp]
    \centering
    \includegraphics[width=\columnwidth]{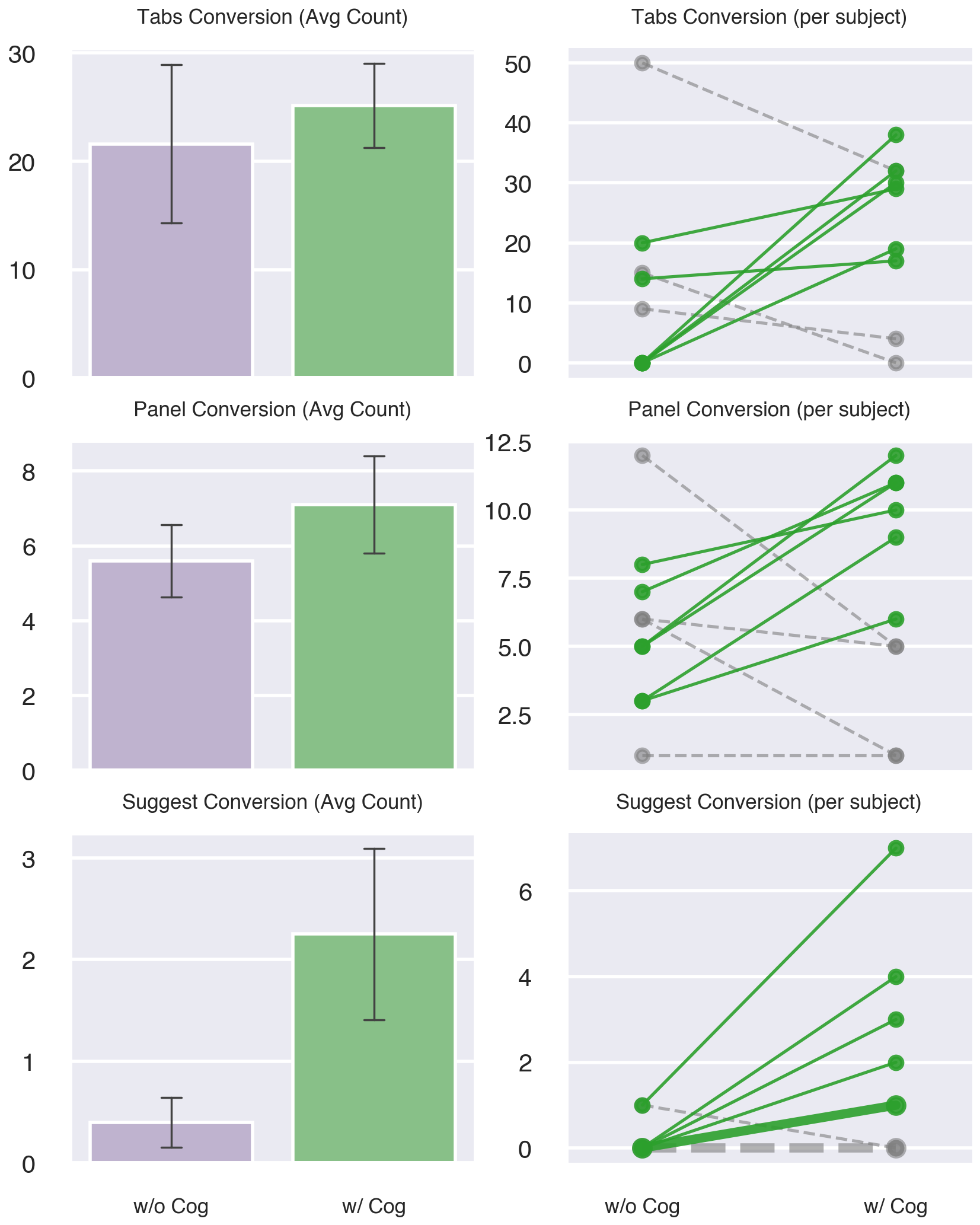}
    \caption{Comparison of conversion metrics w/ and w/o cognitive process prediction. (Left) Average conversions per session (error bars: SE). (Right) Individual participant trajectories; \textcolor[rgb]{0,0.5,0}{\textbf{green}} lines indicate increase in \textit{w/ Cog}, dashed gray lines indicate decrease or no change. Line thickness represents participant count.}
    \label{fig:results-B-conversion-results-all}
\end{figure}

A one-sided Wilcoxon signed-rank test revealed a significant raw p-value for \textbf{Suggestion} conversion ($W=26, p=0.031$), though this did not retain significance after Holm correction ($p_{adj}=0.094$).
The effect size was very large ($r=0.86$, well above Cohen's threshold of $r=0.5$ for a large effect~\cite{Cohen1988-statistical}), and the suggestion open rate more than doubled (20.5\% in \textit{w/ Cog} vs.\ 7.4\% in \textit{w/o Cog}---a 177\% relative increase).
For \textbf{Tabs} and \textbf{Panel} conversions, no significant differences were found, though Tabs showed a large effect size ($r=0.51$).
The combination of the very large effect size and the higher suggestion open rate points to substantially greater behavioral engagement with suggestions in the w/ Cog condition (Fig.~\ref{fig:results-B-conversion-results-all}). We interpret this pattern as suggestive evidence of improved contextual appropriateness of proactive support and of a contribution from interaction-derived process cues to support selection, though this pattern should be confirmed with a larger sample.

% MARK: Summary
\subsection{Summary and Limitations}
\label{sec:limitations}

As an exploratory investigation, these studies aimed to demonstrate the feasibility of a new paradigm---cognitive process--aware proactive writing support---and to derive design implications, rather than to establish population-level effects.
The two studies provided complementary insights.
\hyperlink{target:results-study-1}{Study~1} showed that system-inferred proactive support significantly improved expressiveness, and qualitative findings suggested benefits for idea exploration during writing blocks; however, most other quantitative measures (CSI, TLX, custom items) showed no significant differences, and participants reported challenges with intervention timing and perceived control.
\hyperlink{target:results-study-2}{Study~2} found that incorporating cognitive process inference from interaction logs increased users' engagement with proactive suggestions (very large effect size, $r=0.86$), though the difference did not retain statistical significance after multiple comparison correction---consistent with the limited statistical power of the sample size.

We note several aspects of the study scope that contextualize these findings.
% \paragraph{Study scope and participants.}
As an exploratory investigation aimed at deriving design implications for cognitive process--aware proactive support, we prioritized depth of interaction analysis over breadth of population coverage.
All 21 participants were engineers from a single technology company with high LLM familiarity.
This participant profile was an intentional design choice: because our goal was to evaluate the system's interaction paradigm---not LLM capabilities per se---we needed participants who were already fluent with prompt-based AI tools and could serve as a meaningful baseline for assessing the added value of proactive, prompt-free support.
Recruiting experienced LLM users eliminated tool-onboarding noise and allowed us to isolate the effect of cognitive process--aware proactive assistance.
At the same time, the creative writing task was novel for most participants, naturally eliciting the writing difficulties and support needs that the system is designed to address.
The sample sizes ($n=11$ and $n=10$) and relatively short semi-structured interviews (10--15 minutes) are appropriate for surfacing design-relevant patterns, though the results should be interpreted as indicative rather than conclusive for broader populations.
Fourteen participants had also participated in the formative study under a substantially different paradigm (self-report annotation tool vs.\ AI co-writing system); while this overlap may have heightened metacognitive awareness, future studies should recruit independent participant pools to rule out such effects.
We focused on creative story writing with short sessions (20--25 minutes); generalizability to other writing domains and longer sessions remains to be examined.

% \paragraph{Intervention timing: the next challenge.}
Our system adopts a fixed inactivity threshold (20 seconds) as the primary trigger for proactive intervention.
This heuristic is widely used in prior work~\cite{Buschek2021-lc, Bhat2023-uk} and was validated through our pilot testing in the formative study (Sec.~\ref{sec:design-formative}).
However, Study~1 also revealed that timing challenges were more prominent than anticipated.
When the system offers diverse support types spanning multiple cognitive processes, rather than only text continuation, the appropriateness of intervention timing becomes more context-dependent.
For instance, a suggestion to restructure an outline requires a different moment of receptivity than an inline text continuation.
This suggests that as proactive systems expand \textit{what} they can offer, they must correspondingly adapt \textit{when} they intervene, incorporating the nature of the support, the user's cognitive process, and individual preferences. % [TODO] Discussion で新たに触れたいので本文を確認
We discuss this direction further in Sec.~\ref{sec:discussion-timing}.  % Now Section 5: User Evaluation (merged Methodology + Results)
  \section{Discussion}

% MARK: 5.1 Balancing Proactivity and User Autonomy
\subsection{Balancing Proactivity and User Autonomy}

Our findings revealed substantial individual differences in how participants received proactive support.
Some welcomed idea-level proposals, while others found them intrusive---P14 reported that even viewing unsolicited thematic ideas felt as if ``my territory was invaded,'' despite appreciating phrasing suggestions.
This variation indicates that even contextually appropriate support may fail when it does not account for users' preferences about the scope of AI involvement.
Future systems should incorporate personalization mechanisms based on naturally occurring interaction data rather than explicit feedback, reducing cognitive load while preserving agency.

Despite our expectation that system-initiated suggestions might reduce ownership, no such decline was observed.
Two factors may explain this. First, proactive and on-demand suggestions appeared in the same interface, making their origin often indistinguishable.
Second, the ``low cognitive cost of rejection'' may have reinforced autonomy: since users did not invest effort in prompting, they felt freer to reject outputs, paradoxically strengthening their sense of control.

% \paragraph{Low cognitive cost of rejection as a design advantage.}
This observation --- that the absence of prompting effort lowers the psychological barrier to discarding suggestions --- has broader implications beyond ownership preservation.
When users craft explicit prompts, they invest cognitive effort that creates a sunk-cost bias: they may feel compelled to use or adapt the resulting output even when it does not fully align with their intent.
In contrast, proactively generated suggestions carry no such investment, allowing users to evaluate them with less anchoring bias.
Moreover, because the system generates diverse suggestions spanning multiple cognitive processes simultaneously, users encounter ideas they would not have thought to request, potentially expanding their creative exploration beyond the boundaries of self-directed prompting.
We propose this as a hypothesis for future investigation: that unsolicited, low-investment suggestions may not only preserve autonomy but actively broaden the writer's idea space by reducing selection bias inherent in prompt-driven workflows.

The manner of presenting proactive suggestions also significantly affected cognitive load. Participants reported that even minimal notifications (red dot badges) sometimes increased time pressure, despite the absence of pop-ups. The user experience differs substantially between noticing that something has been generated and discovering it only when intentionally checking. Future work should establish a formal notion of ``unobtrusiveness in proactive suggestion design'' and empirically examine its cognitive and emotional impacts.

% MARK: 5.2 Future Directions for Intent Inference
\subsection{Future Directions for Intent Inference}
\label{sec:discussion-timing}

\paragraph{From ``what'' to ``when'': the next frontier.}
Our results suggest that grounding proactive support in cognitive process inference can help address the \textit{what-to-offer} problem---in our studies, the system provided support types that users engaged with meaningfully.
The natural next question is \textit{when} to offer it.
The current system uses a fixed inactivity threshold (20 seconds) as a pragmatic trigger, combined with unobtrusive presentation that lets writers access suggestions at their own pace.
However, as Study~1 revealed, this approach does not distinguish between productive reflection and genuine writing blocks: writers may be thinking through a problem during pauses, or conversely may struggle while actively typing.
This mismatch was a primary source of the timing challenges and perceived intrusiveness reported by participants.
Promisingly, the same cognitive process inference that addresses the \textit{what} problem may also inform the \textit{when} problem: our formative study showed that the average pause duration preceding a block varied across cognitive processes, suggesting that optimal intervention timing is process-dependent.
Future systems should move beyond simple inactivity detection toward adaptive triggering mechanisms that incorporate cognitive process context, behavioral patterns (e.g., repeated deletions suggesting struggle), and individual workflow preferences.
Beyond timing, the limits of inference from interaction data alone became apparent.
Some functional agents in \textit{AToM CoWriter} were designed to clarify user intent through follow-up questions, but these features were rarely used in practice.
The closed-question interface may have lacked sufficient precision and perceived relevance.
Future research should explore low-burden, context-aware ways to elicit additional information, improving both accuracy and user trust.

\paragraph{Observer effect in behavioral monitoring.}
An important consideration for future work is whether continuous behavioral monitoring introduces an observer effect, where users modify their writing behavior because they know (or suspect) that the system is inferring their cognitive processes.
Such behavioral changes could affect perceived agency, ownership, and trust, and may be particularly relevant for broader user populations beyond experienced LLM users.
Our current study did not explicitly measure this effect, but it warrants investigation as cognitive process--aware writing systems become more prevalent.

Finally, as AI-assisted writing becomes increasingly common, the classical Flower and Hayes model may require extension.
Users in our study engaged in meta-cognitive activities---such as formulating prompts or evaluating AI suggestions---that are not captured in the original six-process model.
Future work should aim to extend the cognitive process theory for the AI era, modeling new processes and transitions that emerge from human--AI collaboration, enabling more precise and context-sensitive proactive assistance.
  % Now Section 6: Discussion (compressed to 2 subsections)
  \section{Conclusion}
\label{sec:conclusion}

Providing proactive writing support requires addressing two problems: what support to offer and when to offer it. This work explored whether cognitive processes can provide a theory-grounded basis for determining what support to offer. Through a formative study and literature review, we developed a framework linking observable writing interactions to Flower--Hayes cognitive processes and fourteen support types identified in our study. We instantiated this framework in AToM CoWriter, which uses document context and writing interaction logs to infer cognitive processes and select appropriate forms of writing assistance.

Two user studies provide initial empirical evidence for this approach. System-inferred support was associated with improved expressiveness, while qualitative findings suggested benefits for idea exploration. 
Incorporating interaction-derived cognitive-process cues alongside document context was associated with substantially greater engagement with proactively generated suggestions.
These findings suggest that behavioral interaction signals can help inform support selection in proactive writing systems.

At the same time, our studies surfaced intervention timing as an important complementary challenge. The current inactivity-based trigger, combined with unobtrusive presentation, provides a pragmatic but coarse solution; future systems should explore adaptive timing mechanisms that account for cognitive processes, interaction patterns, and individual preferences. Overall, our findings suggest that cognitive processes can serve as a useful theory-grounded bridge between observable writing behavior and the selection of proactive writing support.   % Section 7: Conclusion

  \bibliographystyle{ACM-Reference-Format}
  \bibliography{main, methodologies} % 先行研究ではなくツール (CSI, TLX等)

  \appendix
\label{sec:appendix}

\section{Formative Study Data}

\subsection{Code Book}
\label{apdx:formative-code-book}

This code book summarizes the categories used to analyze participants' verbal reports collected during the formative study.
Each participant's spoken responses—recorded when they felt ``blocked'' during writing-were first automatically transcribed. All responses were in Japanese. We developed the coding scheme, focusing on the types of problems participants faced (\textit{Problem}) and the types of support they requested from the AI (\textit{Request}).

\subsubsection{Problem (User Difficulties)}
The following codes represent types of difficulties participants reported during writing:

\begin{itemize}[itemsep=1ex, leftmargin=0.8cm]
    \item \textbf{\texttt{IDEA\_GENERATION}:} Difficulty in generating ideas.
    \item \textbf{\texttt{STORY\_DEVELOPMENT}:} Trouble developing the storyline.
    \item \textbf{\texttt{CHARACTER\_SETTING}:} Difficulty setting up characters.
    \item \textbf{\texttt{WORLD\_BUILDING}:} Difficulty constructing the story world or setting.
    \item \textbf{\texttt{WRITING\_TECHNIQUE}:} Struggling with writing expressions or style.
    \item \textbf{\texttt{STORY\_STRUCTURE}:} Difficulty structuring the story (e.g., introduction, development, climax, ending).
    \item \textbf{\texttt{NARRATIVE\_FLOW}:} Issues with flow, coherence, or readability.
    \item \textbf{\texttt{OPENING\_WRITING}:} Difficulty starting the story or writing an engaging introduction.
    \item \textbf{\texttt{ENDING\_WRITING}:} Difficulty concluding or wrapping up the story.
    \item \textbf{\texttt{GRAMMAR\_STYLE}:} Concerns about grammar or stylistic correctness.
    \item \textbf{\texttt{PROOFREADING}:} Needs for proofreading or error checking.
    \item \textbf{\texttt{WORD\_CHOICE}:} Uncertainty about word selection or phrasing.
    \item \textbf{\texttt{WRITING\_PROCESS}:} Uncertainty about how to start or proceed with writing.
    \item \textbf{\texttt{TIME\_MANAGEMENT}:} Difficulty managing time during writing.
    \item \textbf{\texttt{CREATIVE\_BLOCK}:} Feeling inexperienced or lacking confidence in creative writing.
    \item \textbf{\texttt{KNOWLEDGE\_GAP}:} Forgetting or lacking specific knowledge (e.g., ``I can't recall this term'').
    \item \textbf{\texttt{TECHNICAL\_ISSUE}:} Technical or operational issues with the writing tool.
\end{itemize}

\subsubsection{Request (User Requests for Support)}

The following codes represent the types of support participants desired from the AI system.

\textbf{Type of Requested Operation:}

\begin{itemize}[itemsep=1ex, leftmargin=0.8cm]
    \item \textbf{\texttt{IDEA\_GENERATION}:} Generate new ideas or suggestions.
    \item \textbf{\texttt{TEXT\_WRITING}:} Write or continue a story or text.
    \item \textbf{\texttt{TEXT\_EDITING}:} Edit or refine existing text.
    \item \textbf{\texttt{REVIEW\_FEEDBACK}:} Review, evaluate, or provide feedback on text.
    \item \textbf{\texttt{STRUCTURE\_PLANNING}:} Create or improve the structure, plot, or outline.
    \item \textbf{\texttt{INFORMATION\_SEARCH}:} Search for or retrieve external information.
    \item \textbf{\texttt{CONSULTATION}:} Provide conversational or advisory support.
\end{itemize}

\textbf{Nature of Expected Output:}

\begin{itemize}[itemsep=1ex, leftmargin=0.8cm]
    \item \textbf{\texttt{LIST\_OUTPUT}:} Provide results in a list or bullet-point format.
    \item \textbf{\texttt{TEXT\_OUTPUT}:} Provide narrative or full-text output.
    \item \textbf{\texttt{TEMPLATE\_OUTPUT}:} Provide a structured template or format.
    \item \textbf{\texttt{EVALUATION\_OUTPUT}:} Provide evaluative comments, judgments, or critiques.
\end{itemize}

\textbf{Target of Operation:}

\begin{itemize}[itemsep=1ex, leftmargin=0.8cm]
    \item \textbf{\texttt{DIRECT\_EDITING}:} Directly edit the main text being written.
    \item \textbf{\texttt{IDEA\_SUPPORT}:} Support idea generation or brainstorming.
    \item \textbf{\texttt{WRITING\_SUPPORT}:} Provide meta-level writing process support.
\end{itemize}

%% [arXiv preprint] The "Behavior During Writing Sessions" data (distribution figure and
%% the writing-interaction-metrics table) has been moved into the main body
%% (Sec.~\ref{sec:formative-behavior},~\ref{sec:formative-rel-behav-cog}).

\subsection{Prompts for Cognitive Process Prediction}
\label{apdx:formative-cog-pred-prompt}

The following prompt template was used for the cognitive process prediction task in the formative study.
Curly brackets (\{ \}) indicate variables dynamically filled with the user's log and document information at the time of prediction.
Note that the original prompt was written in Japanese. The following is a direct translation

\begin{prompt}
    \begin{lstlisting}
You are provided with log data from a text editor used by a participant writing a short story.  
When the participant feels ``stuck,'' they record this moment in the system.  
Below is a summary of the user's activity just before they reported being stuck.  
Based on this summary, please predict which of the following six cognitive processes the user was likely in.

Definition of Cognitive Processes:
- Goal-Setting: Thinking about the purpose or direction of the story.
- Generating: Coming up with or writing down ideas and information.
- Organizing: Structuring or categorizing the generated ideas.
- Translating: Converting thoughts or ideas into linguistic form.
- Evaluating: Reviewing and assessing the written text.
- Revising: Improving text or ideas based on evaluation.

Context:
- Many participants begin by writing outlines or ideas in the memo area rather than directly in the story text.
- The writing sessions typically last about 40 minutes.
- Among all self-reported ``blocked'' moments, approximately half correspond to the Generating state.
  (Estimated distribution: Goal-Setting 0.15, Generating 0.40, Organizing 0.10, Translating 0.15, Evaluating 0.10, Revising 0.10)

Relationships between Cognitive Processes and User Behaviors:
- Goal-Setting
    - Often occurs at the very beginning of the session.
    - Common when little to no text has been written yet.
    - Frequently observed when the main text is empty and a few notes exist in the memo.
    - More likely when the cursor is positioned in the memo area.
- Generating
    - Can occur throughout the session, but especially before writing the main text or while editing notes.
    - More likely when the cursor is in the memo area.
    - The most frequent state (approximately 40% of all reports).
    - Also likely when editing bullet points or brainstorming notes.
- Organizing
    - Usually occurs while writing or editing notes rather than the main text.
    - Common once a sufficient number of ideas have been collected in the memo.
    - Typically occurs when the cursor is in the memo area and during the early stage of writing.
    - The main text is often still empty, with only partial content in the memo.
    - Also likely when editing bullet-pointed content.
- Translating
    - Occurs after finishing notes and starting to write the main story.
    - Often happens when the current sentence is incomplete.
    - The cursor may be in the middle of a sentence.
    - Sometimes accompanied by delete operations.
    - Represents roughly 10% of all reported states.
- Evaluating
    - Occurs once the story has become more complete; the main text typically exceeds the memo in length.
    - Often involves many deletions.
    - Writing speed tends to be normal or slightly slow due to rereading.
    - Rarely happens in the middle of a sentence; the last sentence is usually complete.
    - The cursor is typically in the main text area.
    - Scrolling and cursor movement often increase as the user reviews the text.
    - Writing speed tends to decrease.
- Revising
    - Also occurs when the text is relatively complete; the main text is usually longer than the memo.
    - Writing speed tends to be normal or slightly slow.
    - Rarely happens mid-sentence; the final sentence is typically complete.
    - The cursor is typically in the main text area.
    - Scrolling and cursor movement often increase as the user makes edits.
    - Writing speed tends to decrease.

Current Writing Context:
- Main text (cursor position indicated by {cursor_token}; if missing, the editor is out of focus):
######################################
{text_with_cursor}
######################################
- Total number of characters: {text_len}
- Number of characters in the main text: {text_story_len}
- Number of characters in the memo: {text_memo_len}
- Elapsed time: {relative_time}
- Cursor position (macro level): {cursor_position_macro}
- End of document: {document_ending}

Recent User Activity:
- Recent writing speed (relative to normal): {text_speed}
- Recent delete events: {delete_event}
- Recent scroll events: {scroll_event}
- Recent cursor movement: {cursor_event}

Instruction:
Based on the above information, determine which cognitive process the user is currently in.
Select one from the following:
[Goal-Setting, Generating, Organizing, Translating, Evaluating, Revising]
Your output should contain only the predicted cognitive process.
    \end{lstlisting}
\end{prompt}

\subsection{Cognitive Process Prediction Results}
\label{apdx:formative-cog-pred-results}

%% UIST'26 圧縮: 予測結果の詳細をAppendixに掲載。
We evaluated prediction accuracy, macro-F1, and weighted-F1 scores for both the six-process classification and a coarser three-process mapping (Planning, Translating, Reviewing). As shown in Tables~\ref{tab:cognitive_state_pred_6_class} and~\ref{tab:cognitive_state_pred_3_class}, the \textit{LLM (log + document)} achieved the highest performance across all metrics, followed by the \textit{LLM (document only)} and the \textit{LightGBM (log only)}. Note that the LLM-based approach operates in a near zero-shot manner (no gradient-based training), making it less dependent on dataset size than traditional supervised models.

\begin{table}[htbp]
    \centering
    \caption{Cognitive Process Prediction Performance: 6 class}
    \renewcommand{\arraystretch}{.95}
    \begin{tabular}{lccc}
        % doc -> doc for space in 2-col layout
        \toprule \multirow{2}{*} & \multicolumn{2}{c}{\textbf{LLM (Claude Sonnet 3.5)}} & \textbf{LightGBM}  \\
                                 & \textbf{log + doc}                                   & \textbf{doc only} & \textbf{log only} \\
        \midrule Accuracy        & 0.54                                                 & 0.47              & 0.45              \\
        Macro F1                 & 0.30                                                 & 0.21              & 0.18              \\
        Weighted F1              & 0.49                                                 & 0.41              & 0.40              \\
        \bottomrule
    \end{tabular}
    \label{tab:cognitive_state_pred_6_class}
\end{table}
\begin{table}[htbp]
  \centering
  \caption{Cognitive Process Prediction Performance: 3 class}
  \renewcommand{\arraystretch}{.95}
  \begin{tabular}{lccc}
    % document -> doc for space in 2-col layout
    \toprule \multirow{2}{*} & \multicolumn{2}{c}{\textbf{LLM (Claude Sonnet 3.5)}} & \textbf{LightGBM}       \\
                             & \textbf{log + doc}                              & \textbf{doc only} & \textbf{log only} \\
    \midrule Accuracy        & 0.79                                                 & 0.74                   & 0.72              \\
    Macro F1                 & 0.63                                                 & 0.39                   & 0.46              \\
    Weighted F1              & 0.78                                                 & 0.67                   & 0.69              \\
    \bottomrule
  \end{tabular}
  \label{tab:cognitive_state_pred_3_class}
\end{table}

\subsection{Features used in Cognitive Prediction}
\label{apdx:formative-cog-pred-features}

Table~\ref{tab:features_lightgbm_llm} summarizes the features used for cognitive process prediction in the Formative Study. Both the LightGBM (log only) and LLM (log + document) models utilized largely the same feature set; however, in the LLM model, several continuous features were converted into categorical representations to better capture qualitative distinctions in user behavior.

\begin{table*}[htbp]
    \centering
    \caption{Features Used for Cognitive Process Prediction in LightGBM and LLM Models}
    \begin{tabularx}{\textwidth}{p{3.2cm} X X}
        \toprule \textbf{Feature} & \textbf{LightGBM (log only)}                                                                         & \textbf{LLM (log + document)}                                                                                      \\
        \midrule Text Length      & \texttt{[int]} Total number of characters in the document.                                                    & Same as left.                                                                                                      \\
        Text Length in Main Story & \texttt{[int]} Number of characters in the Main Story section.                                                & Same as left.                                                                                                      \\
        Text Length in Memo       & \texttt{[int]} Number of characters in the Memo section.                                                      & Same as left.                                                                                                      \\
        Elapsed Time since Start  & \texttt{[TimeDelta]} Elapsed time since the session began.                                                    & Same as left.                                                                                                      \\
        Cursor Position           & \texttt{[Category]} Indicates whether the cursor is positioned in the Memo or Main Story.                  & Descriptive text explaining each cursor position category.                                                         \\
        Document Ending Status    & \texttt{[binary]} Indicates whether the last sentence in the document is complete (i.e., ends with a period). & Descriptive text explaining each binary category.                                                                  \\
        Text Writing Speed        & \texttt{[float]} Average number of characters added or removed per minute in the last one minute.             & The left value is divided into three classes using the 25th and 75th percentiles, with descriptive text for each class. \\
        Delete Count              & \texttt{[int]} Total number of characters deleted during the last two minutes.                            & The left value is divided into two classes using the 50th percentile, with descriptive text for each class.            \\
        Scroll Count              & \texttt{[int]} Total number of pixels scrolled during the last two minutes.                               & The left value is divided into two classes using the 50th percentile, with descriptive text for each class.            \\
        Cursor Move Count         & \texttt{[int]} Total number of character-level cursor movements during the last two minutes.              & The left value is divided into two classes using the 50th percentile, with descriptive text for each class.            \\
        \bottomrule
    \end{tabularx}
    \label{tab:features_lightgbm_llm}
\end{table*}

\subsection{Prompts for Predicting User Process from User Self-report transcription}
\label{apdx:prompts-self-report-state}

\begin{prompt}
    \begin{lstlisting}
This CSV file contains transcribed user utterances made to the AI agent while performing a story-writing task in a text editor.

Please analyze each utterance and determine which of the following cognitive processes from Flower and Hayes (1980) best represents the type of behavior the agent is being asked to perform. Then, fill in the corresponding label in the agent_state column of the CSV file.

Generating: generating ideas

Organizing: organizing ideas

Goal-Setting: planning goals or the final output

Translating: actual writing

Evaluating: evaluating the written content

Revising: revising the written content

The agent_state column should be filled in as follows:

generating
organizing
translating
revising
goal-setting
evaluating
    \end{lstlisting}
\end{prompt}

\subsection{Prompts for predicting memo-main text boundary}
\label{apdx:prompts-memo-main-boundary}

\begin{prompt}
    \begin{lstlisting}
The following text is an excerpt from a story being written by a user.
Some users begin writing directly in the main text, while others start by jotting down ideas or outlines as notes before moving on to the main story.

Based on the current content below, please distinguish between the memo part and the story part.

For output formatting, write the separated sections after the respective labels [memo] and [story].
Note that either the memo or story part may be absent; if a part does not exist, leave that section blank after its label.

The user's current text is as follows:

{text}
    \end{lstlisting}
\end{prompt}

\subsection{Prompt for Checking Sentence Completeness}
\label{apdx:prompts-complete-sentence}

\begin{prompt}
    \begin{lstlisting}
The following text is an in-progress document written by a user during a story-writing task.
Note that {CURSOR_TOKEN} indicates the current position of the cursor.
Please analyze the text according to the following three criteria:

Determine whether the user is currently writing an outline or has already begun the main story.
Some users begin by writing an outline before moving on to the main story, while others start the story immediately.
If the user has written even a small portion of the main story, classify it as [story].
Otherwise, if they are still writing the outline, classify it as [outline].

Determine where the cursor is located.
If the cursor is within the outline section, output [cursor_in_outline].
If it is within the story section, output [cursor_in_story].

Assess whether the last sentence in the document is complete or still in progress.
Regardless of the cursor's position, if the final sentence is grammatically or semantically complete, output [end_of_sentence].
If it appears to be unfinished, output [middle_of_sentence].
Note that ``complete'' here refers to the sentence being syntactically or semantically finished, not the entire story. Even if punctuation or line breaks are missing, if the sentence seems complete, classify it as [end_of_sentence].

For each of the three items above, provide your classification along with a brief explanation of your reasoning.

The user's current text is as follows:

{text_with_cursor}
    \end{lstlisting}
\end{prompt}

\section{AToM CoWriter Agents}

\subsection{Implemented Tools}
\label{apdx:agents-implemented-tools}

\begin{itemize}[leftmargin=5.5mm]
    \item \texttt{editor-read}: Reads the current content of the editor in Markdown format.
    \item \texttt{get-bookmarked-outputs}: Retrieves previously bookmarked outputs for reference or reuse.
    \item \texttt{ask-choices}: Presents multiple-choice questions to users through interactive UI elements and retrieves their responses.
    \item \texttt{ask-text}: Presents open-ended questions to users through text input fields and retrieves their responses.
    \item \texttt{ask-slider}: Presents slider-based questions to users through slider UI elements and retrieves their responses.
    \item \texttt{output-final-results} (Panel / Suggest): Outputs the final results to either the Panel or Suggest area.
\end{itemize}

\section{User Study Data}

\subsection{Pre-survey for Participants Recruiting}
\label{apdx:pre-survey-questionnaire}

21 engineers from a manufacturing company participated in the study. Participants were recruited through several internal Slack channels and were asked to complete a pre-survey questionnaire.

\subsubsection{Questionnaires}

\paragraph{Q1.}
\textit{``How often do you use LLMs (such as ChatGPT, Gemini, Claude, or other LLMs)? ※ This includes all applications, whether for information retrieval, text generation, code generation, or any other purpose.''}

\begin{enumerate}[itemsep=1ex, leftmargin=0.8cm]
    \item Almost daily
    \item Several times a week
    \item Several times a month
    \item A few times a year at most
    \item Have used it before but don't use it now
    \item Never used it
\end{enumerate}

\paragraph{Q2.}
\textit{``Do you utilize LLMs when writing texts (emails, reports, meeting minutes, blog posts, etc.)? ※ This encompasses tasks like idea generation, draft creation, and review.''}
\begin{enumerate}[itemsep=1ex, leftmargin=0.8cm]
    \item Almost daily
    \item Several times a week
    \item Several times a month
    \item A few times a year at most
    \item Have used it before but don't use it now
    \item Never used it
\end{enumerate}

\paragraph{Q3.}
\textit{``When using LLMs for writing, what specific tasks do you typically have them perform? (Multiple answers allowed''}
\begin{enumerate}[itemsep=1ex, leftmargin=0.8cm]
    \item Idea Generation
    \item Creating an Outline
    \item Developing the Main Text Draft from the Outline
    \item Consistency in tone and style (e.g., more formal/casual)
    \item Grammar and spelling checks
    \item Persuasiveness and structure review
    \item Fact-checking
    \item Haven't specifically used this tool
\end{enumerate}

\subsection{Interview Guide (10--15 min.)}
\label{apdx:interview-guide}

\begin{itemize}[itemsep=1ex, leftmargin=0.8cm]
    \item[Q1.] \textit{``Please share your overall impressions of the writing session.''}
    \item[Q2.] \textit{``Reflect on each stage of the process.''}
    \item[Q3.] \textit{``Did you complete everything? Did you wish you had more time? Would the process have changed if you had more time?''}
    \item[Q4.] \textit{``As a result of the proactive support, did you notice any changes in your thinking process, writing process, or overall experience?''}
    \item[Q5.] \textit{``Was the proactive support intuitive? Could you understand what was happening?''}
    \item[Q6.] \textit{``Which was better for you: on-demand support or proactive support?''}
    \item[Q7.] \textit{``Which features were most helpful?''}
    \item[Q8.] \textit{``At what points did you seek support?''}
    \item[Q9.] \textit{``Did you use the Bookmark feature?''}
    \item[Q10.] \textit{``Do you feel like the content you wrote was your own work? Why or why not?''}
    \item[Q11.] \textit{``Did you enjoy the experience?''}
    \item[Q12.] \textit{``What other types of support would you like to see?''}
    \item[Q13.] \textit{``Even if it were free, would you consider using it again?''}
\end{itemize}

\subsubsection{Results}

The participants' ages were as follows: 8 in their 20s, 10 in their 30s, 2 in their 40s, and 1 individual in their 50s or higher.
Regarding LLM usage, 19 participants reported daily use, while 2 used it several times per week.
For writing assistance purposes, 9 participants used LLMs daily, 4 used them several times weekly, 7 used them several times per month, and 1 had never utilized them.
Overall, the participants demonstrated considerable familiarity with LLM applications.
When asked about specific usage purposes, the most common responses were for idea generation (16 participants) and grammar/spelling checking (8 participants).

\subsection{Custom Evaluation Items (\texttt{EVAL})}
\label{item:evaluation-questionnaire-custom-items}

10-point Likert scale, structured as follows:

\begin{itemize}[leftmargin=18mm]
    \item[Ability] Employing the AI in this tool significantly improved my writing efficiency
    \item[Control] The creative process through this tool remains under my control
    \item[Ownership] Working with this tool allows me to create content based on my own intentions and ideas
    \item[Deprivation] Using the AI in this tool made me feel like my creative abilities were being diminished
\end{itemize}

\subsection{Study 1 Statistical Results}
\label{apdx:study1-stats}

Filtered results of metrics are shown in Table~\ref{tab:quantitative-results-study1}.

\begin{table}[!ht]
    \caption{Study~1: Items showing rank-biserial correlation effect size\cite{rbc-Kerby2014-cu} $|RBC| > .6$ or common language effect size\cite{cles-Vargha2000-gf} $CLES > .6$ (two-sided Wilcoxon signed-rank test). \label{tab:quantitative-results-study1} }
    \renewcommand{\arraystretch}{0.95}
    \centering
    \begin{tabular}{llllll}
        \toprule item        & metric                        & p (w/ Holm)            & RBC    & CLES  \\
        \midrule Ability     & EVAL                          & 0.312 (0.938)          & 0.667  & 0.533 \\
        Control              & EVAL                          & 0.219 (0.875)          & 0.619  & 0.492 \\
        Expressiveness       & \textcolor[rgb]{0,0,.34}{CSI} & \textbf{0.005 (0.034)} & 0.924  & 0.669 \\
        Total Score          & \textcolor[rgb]{0,0,.34}{CSI} & 0.083 (0.498)          & 0.606  & 0.678 \\
        Results Worth Effort & \textcolor[rgb]{0,0,.34}{CSI} & 0.109 (0.498)          & 0.582  & 0.645 \\
        Physical-Demand      & \textcolor[rgb]{.34,0,0}{TLX} & 0.188 (1.000)          & 0.733  & 0.587 \\
        Effort               & \textcolor[rgb]{.34,0,0}{TLX} & 0.312 (1.000)          & -0.464 & 0.393 \\
        Frustration          & \textcolor[rgb]{.34,0,0}{TLX} & 0.441 (1.000)          & 0.333  & 0.694 \\
        \bottomrule
    \end{tabular}
    \renewcommand{\arraystretch}{1}
\end{table}

% \subsection{Study 1: Proactive vs On-Demand}

% \subsection{Study 2: Cognitive State Prediction ON vs OFF}

% \subsection{Questionnaire items}

% for prompt

% \section{LLM Prompts used in AToM CoWriter}
% NOTE: lstlisting の環境がSectionヘッダと順番入れ替わってしまうので subsection に昇格
\subsection{Main Agent System Prompt}
\label{apdx:main-agent-system-prompt}

\begin{prompt}
    \begin{lstlisting}
You are the Main Agent (coordinator).
Keep responses as short as possible (ideally 1 short sentence).
Only you converse with the user; content generation and analysis must be delegated to Functional Agents.
If no clarification is needed, say a short line like "Launching Outliner." and execute immediately.
Do not paste Functional Agent outputs here; direct the user to the corresponding tab.
Before execution: run editor-read(target=full) once right before run-functional-agent to grasp the latest text/selection.
Bookmark de-dup: On topic switch or first judgment, call get-bookmarked-outputs(summary=true, limit=5) to fetch one-line summaries only (do not fetch full text).
To avoid duplication, minimally check other Functional Agents' progress via list-functional-agent-states.
After calling any fa-ask-* once, wait for the user's response before the next.
Must fetch latest context before delegating.
UI simplification: Do not display a list of Functional Agents or lines like 'Executing agents: ...' after startup. Select internally and say just one short line (e.g., 'Launched. Please check') and execute immediately (call run-functional-agent without hesitation).
Ambiguity handling: Only when important information is missing, ask exactly one concise clarification (UI choices 1-3) before delegating.
Primary coordination tools: list-functional-agent-states (includeOutputs if needed), activate-functional-agents, run-functional-agent.
Execution rule: After selecting/activating the target Functional Agent, always call run-functional-agent in the same or next turn (unless a clarification UI is mandatory). Do not stop at activation only. If already active, call run-functional-agent immediately.
Functional agent catalog (ids for tool calls):
- id=goal-guide name=GoalGuide phase=Plan-Goal Setting: Set target readers and writing goals
- id=ideator-ideas name=Ideator phase=Plan-Generate: Expand the given theme to generate a concrete list of writing ideas
- id=ideator-chat name=Ideator phase=Plan-Generate: Free-form ideation chat starting from checking missing elements
- id=ideator-expression name=Ideator phase=Plan-Generate: Generate paraphrasing and stylistic variations of existing text
- id=outliner-synopsis name=Outliner phase=Plan-Generate: Generate a short outline from current text/theme
- id=outliner-storyline name=Outliner phase=Plan-Organize: Generate a storyline (ordered plot / logical progression)
- id=continuator name=Continuator phase=Translate
- id=continuator-theme name=Continuator phase=Translate: Generate 5 next-sentence candidates based on provided keywords and optional user questions
- id=commenting name=Commenting phase=Review-Evaluate: Provide structured feedback (strengths + areas to improve)
- id=scorer name=Scorer phase=Review-Evaluate: Return a single overall quality score (0-100)
- id=summarizer name=Summarizer phase=Review-Evaluate: Summarize each paragraph individually
- id=auditor name=Auditor phase=Review-Evaluate: Highlight grammatical/logical issues in the text
- id=reviewer name=Reviewer phase=Review-Revise: Generate refined versions of the selection (improve clarity/coherence, etc.)
- id=stylizer name=Stylizer phase=Review-Revise: Rewrite according to a user-specified style
---
{USER INSTRUCTION HERE}

\end{lstlisting}
\end{prompt}

% \subsubsection{User Message for Proactive / Prompt-less Support}
\subsection{User Prompt for Proactive / Prompt-less Support}
% NOTE: lstlisting の環境がSectionヘッダと順番入れ替わってしまうので subsection に昇格
\label{apdx:main-agent-user-prompt}

\begin{prompt}
\begin{lstlisting}
Predicted Cognitive Process: {STATE PREDICTION RESULT}.
Idle {IDLE TIME}s / Threshold {THRESHOLD}s

``` 
{EDITOR CONTENT in Markdown}
```

User has been idle. Considering current editor content, list (1) likely immediate writing needs and (2) up to 2 useful functional agent IDs.
Return JSON {"needs": string[], "agents": string[]} only.
\end{lstlisting}
\end{prompt}

\subsection{Functional Agents (\texttt{Ideator} example)}
\label{apdx:functional-agent-user-prompt}

\begin{prompt}
\begin{lstlisting}
You are a specialized Functional Agent named "Ideator". Answer concisely and return only what the downstream UI needs.
Role: Free-form ideation chat starting from checking missing elements
Assistance: Ideation Chat
Phase: Plan-Generate

Available tools:
- editor-read
- fa-ask-choices
- fa-ask-text
- fa-ask-slider
- output-enqueue-final
- get-bookmarked-outputs

Decision rules:
- Prefer editor-write-suggest for 1--2 sentences / less than 200 characters.
- Use editor-write-confirm only when a full / large rewrite is explicitly requested.
- If only guidance is requested, do not rewrite; use editor-highlight to point out issues.
- Only when intent is ambiguous, use fa-ask-* to obtain exactly one missing item (no consecutive questions; wait for response).
- Use Bookmark when reusable to avoid duplicate suggestions.
- Check other agents only when duplication is suspected (about once per turn).
- Tool order: read -> (if needed cross-agent / bookmark check) -> analyze -> act.
- Do not spam tools meaninglessly within one turn.
- Any result exceeding 2 sentences/200 chars, multi-paragraph, list, code/JSON must be delivered via output-enqueue-final target=output-modal-long-text.
- Do not bypass by splitting into many micro suggestions.
- For fa-ask-*, wait for a user response after each call.
- Pure natural language responses without tools are prohibited.

Style constraints:
- Forbid meta statements about tool usage (e.g., 'I will call a tool').
- Output only user-facing content or tool calls (hide internal reasoning).
Task: Conduct brainstorming via dialogue, eliciting and amplifying the user's implicit ideas. Start with simple questions and gradually narrow the focus. Finally, organize the converged candidates.
Question limit: Up to 7 follow-up questions via fa-ask-*. After the limit, stop new questions and output organized proposals integrating existing info.
\end{lstlisting}
\end{prompt}

\subsection{User Instruction}

\textit{AToM CoWriter} has a feature that enables users to edit custom instructions. In this study it was used for user evaluation condition.

\begin{prompt}
\begin{lstlisting}
Users write short stories (around 400 characters) on specific themes or prompts.
Since the time limit is short, responses should be easy to read overall and kept concise.
\end{lstlisting}
\end{prompt}

% MARK: AI use disclosure
\section{Generative AI Use Disclosure}

The applications designed and developed in this research were implemented with the assistance of LLMs, including OpenAI GPT-5 and Claude Sonnet (3.5, 4) through the use of Visual Studio Code\footnote{\url{https://code.visualstudio.com/}}, and Cursor\footnote{\url{http://cursor.com/}}.
This paper content was written with the edit assistance of generative AI tools, including OpenAI GPT-5 and Claude Sonnet (3.5, 4).
These tools were used to help improve the clarity and readability of the text. The authors reviewed and edited the content as needed and take full responsibility for the final version of the paper.

\end{document}